\documentclass[12pt,preprint]{aastex}
\usepackage{graphicx}
\slugcomment{Accepted for Publication in the Astrophysical Journal}

\begin{document}
\title{A Deep {\em Chandra} ACIS Study of NGC 4151. III. the Line Emission and Spectral Analysis of the Ionization Cone}

\author{Junfeng Wang\altaffilmark{1}, Giuseppina
  Fabbiano\altaffilmark{1}, Martin Elvis\altaffilmark{1}, Guido
  Risaliti\altaffilmark{1,2}, Margarita Karovska\altaffilmark{1},
  Andreas Zezas\altaffilmark{1,3}, Carole G. Mundell\altaffilmark{4},
  Gaelle Dumas\altaffilmark{5}, and Eva Schinnerer\altaffilmark{5}}

 \altaffiltext{1}{Harvard-Smithsonian Center for Astrophysics, 60 Garden St, Cambridge, MA 02138}
 \altaffiltext{2}{INAF-Arcetri Observatory, Largo E, Fermi 5, I-50125 Firenze, Italy}
 \altaffiltext{3}{Physics Department, University of Crete, P.O. Box 2208, GR-710 03, Heraklion, Crete, Greece}
 \altaffiltext{4}{Astrophysics Research Institute, Liverpool John Moores University, Birkenhead CH41 1LD, UK}
 \altaffiltext{5}{Max-Planck-Institut f$\ddot{\rm u}$r Astronomie, K$\ddot{\rm o}$nigstuhl 17, D-69117 Heidelberg, Germany}

\email{juwang@cfa.harvard.edu}

\begin{abstract}

This paper is the third in a series in which we present deep Chandra
ACIS-S imaging spectroscopy of the Seyfert 1 galaxy NGC 4151, devoted
to study its complex circum-nuclear X-ray emission.  Emission features
in the soft X-ray spectrum of the bright extended emission
($L_{0.3-2{\rm keV}}\sim 10^{40}$ erg s$^{-1}$) at $r>130$ pc
(2\arcsec) are consistent with blended brighter OVII, OVIII, and NeIX
lines seen in the {\em Chandra} HETGS and XMM-Newton RGS spectra below
2 keV.  We construct emission line images of these features and find
good morphological correlations with the narrow line region clouds
mapped in [OIII]$\lambda$5007\AA.  Self-consistent photoionization
models provide good descriptions of the spectra of the large scale
emission, as well as resolved structures, supporting the dominant role
of nuclear photoionization, although displacement of optical and X-ray
features implies a more complex medium.  Collisionally ionized
emission is estimated to be $\la$12\% of the extended
emission. Presence of both low and high ionization spectral components
and extended emission in the X-ray image perpendicular to the bicone
indicates leakage of nuclear ionization, likely filtered through warm
absorbers, instead of being blocked by a continuous obscuring torus.
The ratios of [OIII]/soft X-ray flux are approximately constant
($\sim$15) for the 1.5~kpc radius spanned by these measurements,
indicating similar relative contributions from the low- and
high-ionization gas phases at different radial distances from the
nucleus.  If the [OIII] and X-ray emission arise from a single
photoionized medium, this further implies an outflow with a wind-like
density profile.  Using spatially resolved X-ray features, we estimate that
the mass outflow rate in NGC 4151 is $\sim$2$M_{\odot}$ yr$^{-1}$ at
130 pc and the kinematic power of the ionized outflow is $1.7\times
10^{41}$ erg s$^{-1}$, approximately $0.3\%$ of the bolometric
luminosity of the active nucleus in NGC 4151.

\end{abstract}


\section{Introduction}

The Seyfert 1 galaxy NGC 4151 is well known to have an extended narrow
line region (ENLR) in the optical (e.g., Perez et al. 1989; Robinson
et al. 1994; Crenshaw et al. 2000), indicating interaction between the
active galactic nucleus (AGN) and the host galaxy (Unger et al. 1987).
Thanks to its proximity ($d\sim 13.3$ Mpc, $1\arcsec=65$ pc; Mundell
et al. 1999) and extensive coverage at all wavelengths (Ulrich 2000
and references therein), NGC 4151 thus represents an excellent
laboratory to study observationally nuclear feedback.

Presence of extended emission in the soft X-rays is also well
established (Elvis et al.\ 1983; Morse et al.\ 1995; Ogle et
al.\ 2000; Yang et al.\ 2001; Wang et al. 2010a,2011a).  The nature of
the kpc-scale extended X-ray emission remains uncertain; suggestions
include collisionally ionized plasma, photoionized gas, a hybrid of
both, or electron-scattered nuclear emission (Elvis et al. 1983, 1990;
Heckman \& Balick 1983; Morse et al. 1995; Schulz \& Komossa 1993;
Ogle et al. 2000; Komossa 2001; Yang et al. 2001), although
electron-scattering is less favored, because the required scattering
column and spectral variability do not match the observations (Weaver
et al. 1994; Yang et al.\ 2001). Previous {\em Chandra} imaging
studies of NGC 4151 (Ogle et al. 2000; Yang et al. 2001; Gonzalez
Martin 2008; Wang et al. 2009, 2010a) found good morphological
agreements between the extended X-ray emission and the optical
forbidden line emission (in particular [OIII] $\lambda$5007).  Such a
close spatial correspondence between the soft X-rays and the [OIII]
emission appears to be common in nearby Seyfert 2 galaxies, leading
Bianchi et al.\ (2006) to propose that optical and X-ray features
arise from a single photoionized medium.

High spectral resolution grating observations of NGC 4151 with {\em
  Chandra} and {\em XMM}-Newton further show that the soft X-ray
emission is due to both blended emission lines and radiative
recombination continua (RRC) from He-like and H-like transitions of
carbon, oxygen, neon, and nitrogen (Ogle et al. 2000; Schurch et
al. 2004; Armentrout et al. 2007).  The X-ray line ratio diagnostics
and electron temperatures of the RRC features support the conclusion
that most of the gas is photoionized ($T\sim 10^4$ K) by the AGN
(Schurch et al. 2004; Armentrout et al. 2007), although
photoexcitation by the AGN continuum may also play an important role
(Kinkhabwalla et al. 2003, Ogle et al. 2003).

However, these grating studies of NGC 4151 cannot provide the spatial
distribution of each spectral feature.  The extracted spectra consist
of both nuclear emission and extended emission from various physical
scales.  Therefore, based on the high resolution spectra alone, one
cannot exclude that the photoionized emission may be associated with
the bright narrow-line region (NLR) gas clouds close to the nucleus
($\ll 100$~pc), while a significant contribution from collisionally
ionized gas may be present at larger radii (e.g., NGC 1365, Wang et
al.\ 2009, Guainazzi et al.\ 2009).  Although Yang et al. (2001)
presented ACIS imaging data, the limited signal to noise ratio ($S/N$)
of these data prevented detailed high spatial resolution comparisons
of optical [OIII] and X-ray morphology, or the spectral analysis of X-ray
emission from the spatially resolved features.

The X-ray emission mechanisms in this prototypical Seyfert 1 galaxy
deserve further investigation with high quality X-ray data that allow
spatially resolved spectral analysis of the circum-nuclear region.
This is the aim of the study presented in this paper, the sixth and
last in our series of studies on the nuclear and circum-nuclear region
of NGC 4151 using deep {\em Chandra} ACIS and HRC observations (PI:
Fabbiano).  Using the HRC data and image deconvolution techniques,
Wang et al.\ (2009) resolved the nuclear emission on spatial scale of
$\sim$30 pc and made comparisons with {\em HST} observations.  Wang et
al.\ (2010a) presented the X-ray spectral analysis of the NGC 4151
nuclear emission.  Wang et al. (2010b) reported the discovery of faint
soft X-ray emission extending 2 kpc out from the active nucleus and
filling in the cavity of the H I material.  In Paper I of the present
series (Wang et al.\ 2011a), based on in-depth analysis of the ACIS
data set, we have examined the kpc scale X-ray morphology and the
relations between X-ray absorption features and the cold ISM in the
host galaxy.  In Paper II (Wang et al.\ 2011b) we have reported on
strong evidence of jet--cloud interaction in the nuclear region
($r\leq 130$ pc) and compared these findings to our previous HRC
study.  Here we focus on the comparison between the X-ray line
emission of the ionization cone and the ionized gas traced by the
optical line emission, and we report on the spectral analysis of
spatially resolved X-ray features.  The paper is organized as
follows. In \S~2, we first briefly describe the data reduction,
present the X-ray emission line morphologies (\S~2.1) and the spectral
modeling of all extended emission within a $30\arcsec$-radius ($\sim
2$~kpc) and the spatially resolved emission (\S~2.2).  The results are
discussed in \S~3 and summarized in \S~4.

\section{Data Extraction and Analysis}

NGC 4151 was observed by {\em Chandra} for a total of 180 ks (after
screening for high background intervals) with the spectroscopic array
of the Advanced CCD Imaging Spectrometer (ACIS-S; Garmire et
al.\ 2003) in 1/8 sub-array mode during March 27-29, 2008. We have
presented the details of our ACIS observations of NGC 4151 and the data
reduction in Paper I.

The data were analyzed following the standard procedures using CIAO
(Version 4.2) with the CALDB 4.2.1 provided by the {\em Chandra} X-ray
Center (CXC).  Subpixel event repositioning and subpixel binning
techniques (Paper I and references therein) have been applied to the
ACIS images to improve the spatial resolution.

The complexity in the data analysis caused by the bright nuclear
emission was described in Papers I and II.  We established that for
the $soft$ X-ray emission ($E<1$~keV), photon pile-up is most severe
in the inner $r<1\arcsec$, mild at $1\arcsec \la r \la 2 \arcsec$, and
not an issue at $r > 2 \arcsec$.  {\em Chandra} PSF simulations were
performed to provide an estimate of the expected contamination from
the nuclear emission in an extended feature.  We have taken into
account this information in the following analysis.

\subsection{Broad-band Images and X-ray Emission Line Maps}

\begin{figure}
\epsscale{0.6}
%\plotone{fig1_1223.ps}
%\plotone{circ_profile2.ps}
\plotone{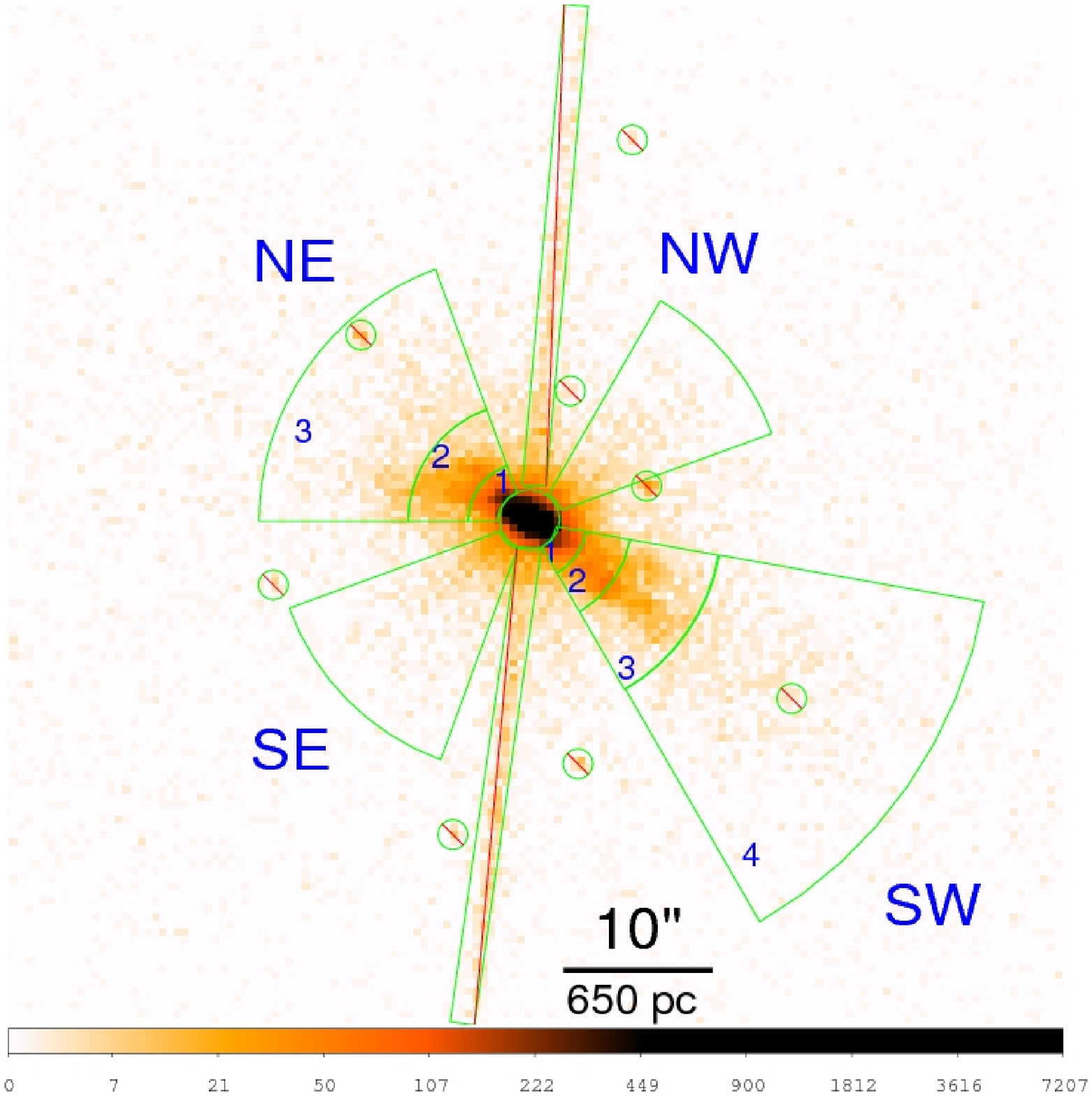}
\plotone{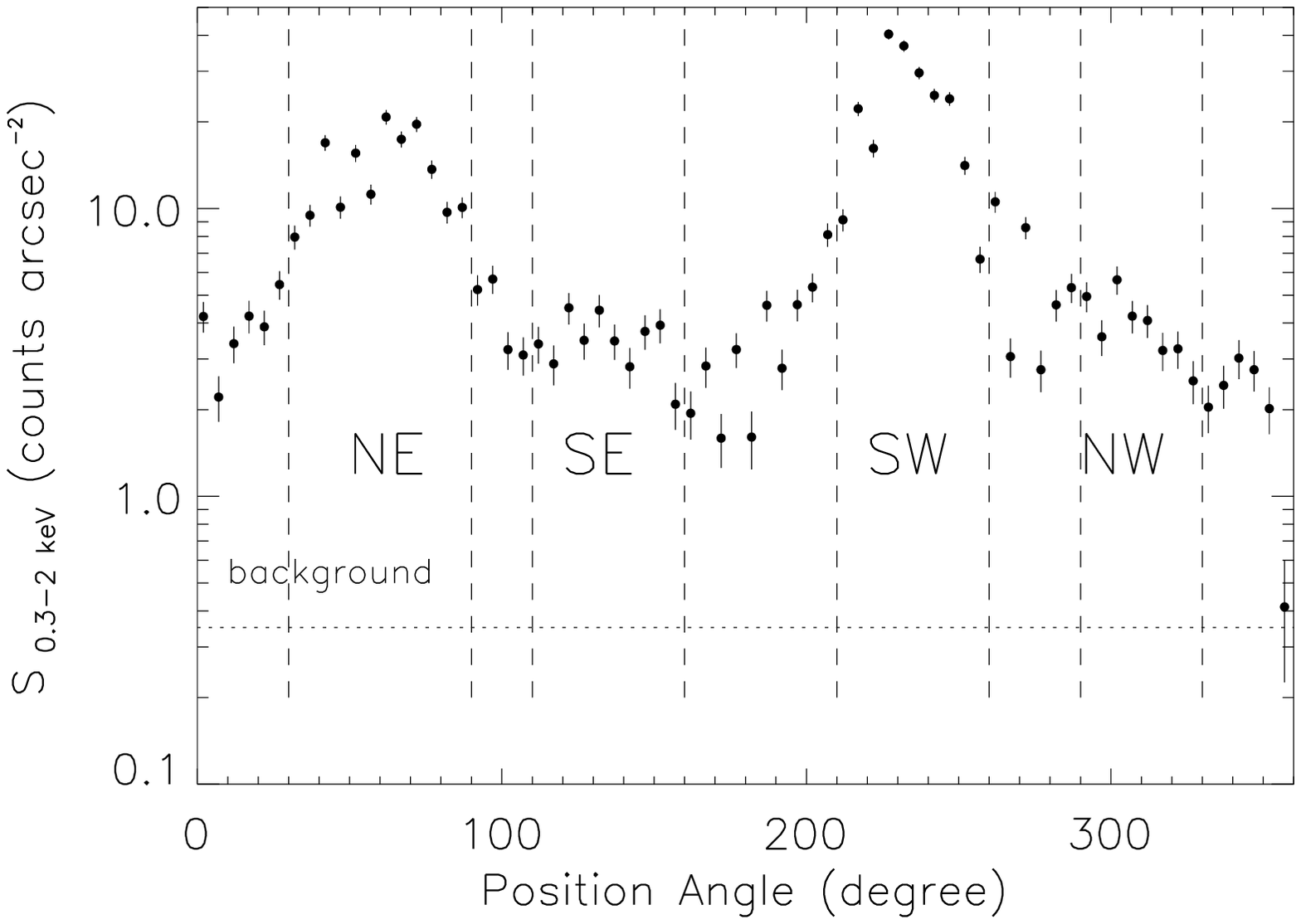}
\caption{(a) Extraction regions of the large scale extended emission
  superposed on a gray-scale ACIS 0.3--2 keV band image. Excluded
  point sources are indicated by circles. The readout streaks running in the north-south direction is indicated. (b) The azimuthal surface
  brightness profile of the 0.3--2 keV emission, outlining sectors
  that contain bright extended emission.
\label{define}}
\end{figure}

\begin{figure}
\epsscale{0.8}
%\plotone{smoo_str_a.ps}
\plotone{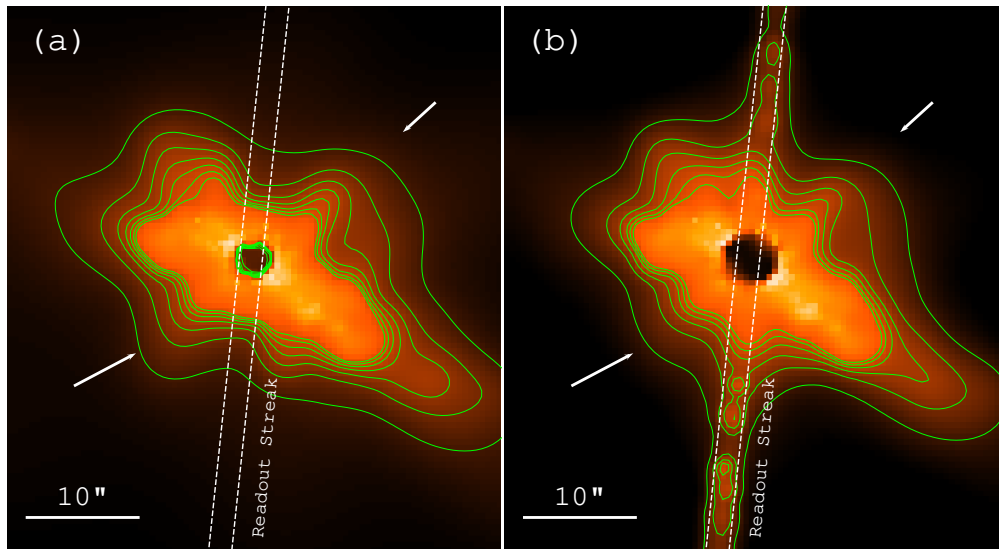}
\caption{(a) Adaptively smoothed ACIS image (0.3--2 keV) of the
  central 3 kpc-diameter region of NGC 4151.  (b) Same as (a), but
  here the readout streaks are not removed, in order to avoid
  artificially increasing the significance of the NW--SE extension
  (indicated by the arrows).  The contours are overlaid to help
  visualize the extent of the diffuse emission.  The brightest nuclear
  region is masked.
\label{contour}}
\end{figure}

\begin{figure}
\epsscale{0.7}
%\plotone{3profile.ps}
%\plotone{3profile_b.ps}
\plotone{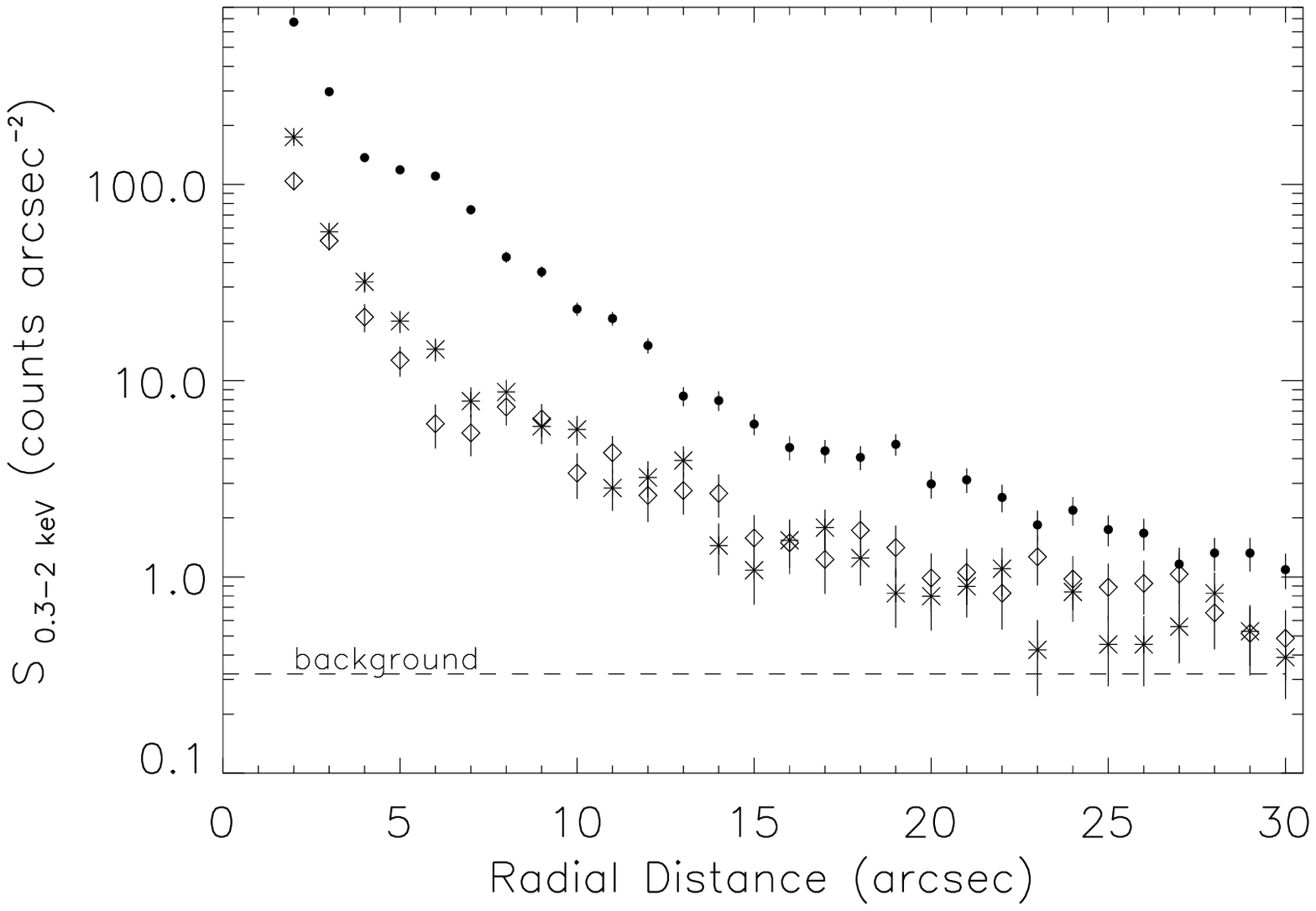}
\plotone{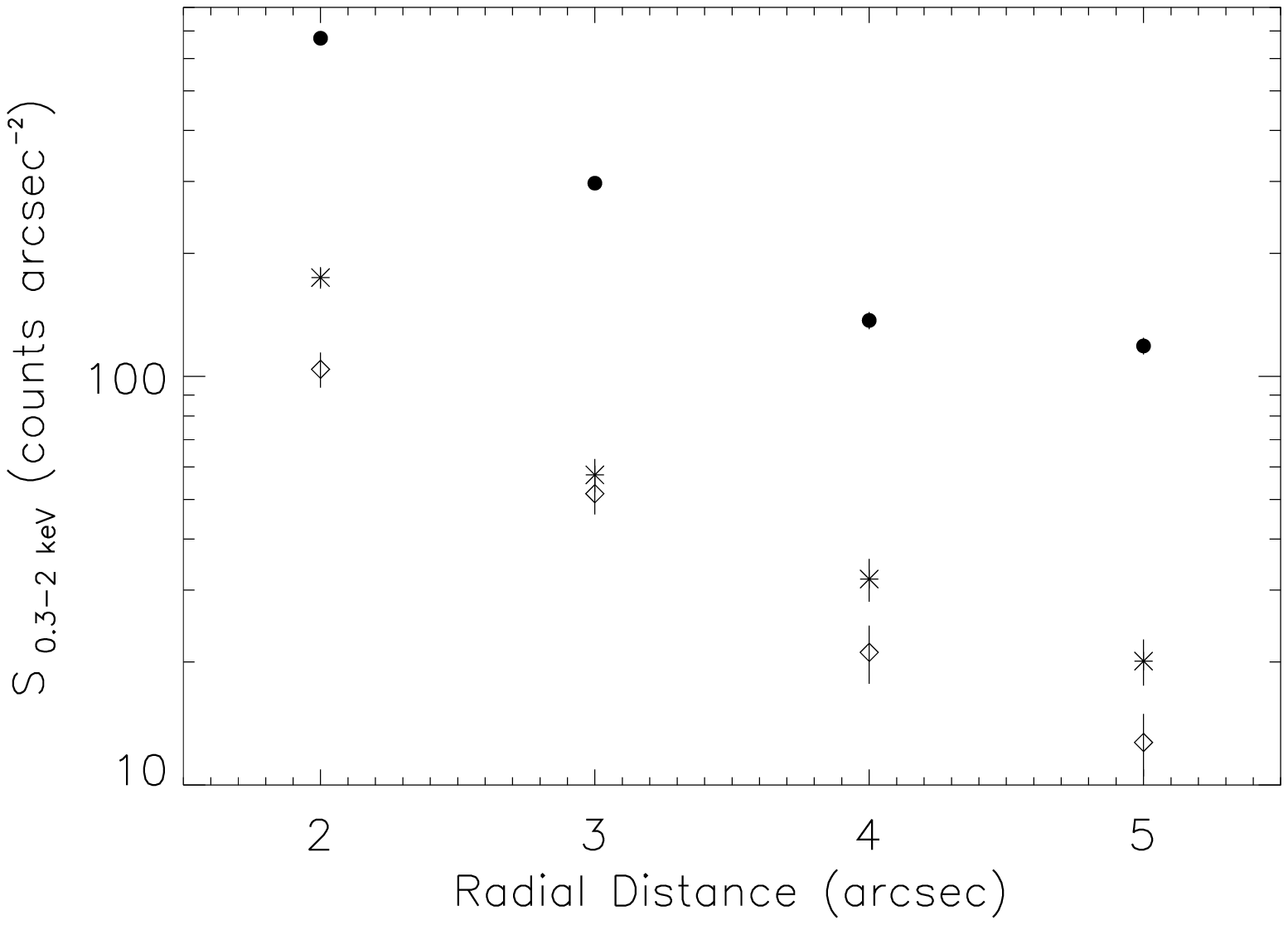}
\caption{(a) Comparison of the radial profiles of the bright SW cone
  (dots), the NW sector (asterisk), and the region in between (along
  $P.A.\sim 270^{\circ}$; diamond). (b) Same as (a) but emphasizing
  the differences in the inner 5 arcsec region.
\label{3profile}}
\end{figure}

Figure~\ref{define} (top) shows the 0.3--2 keV ACIS image
($\sim$0.5\arcsec\/ per pixel). Bright structured soft X-ray emission
along the northeast (NE) -- southwest (SW) direction is apparent,
together with fainter, less extended emission along the northwest (NW)
-- southeast (SE) direction.  The inner $r<2 \arcsec$ nuclear region,
where we identified interactions between the radio jet and the optical
clouds, has been discussed in Paper II. Here we focus on the kpc-scale
extended emission.  We first extracted counts from every $10^{\circ}$
sectors to create azimuthal surface brightness profiles for the
extended X-ray emission ($2\arcsec \leq r \leq 30 \arcsec$), which
provides a clear view of its angular distribution, and to identify
sectors that contain the bright outflows (Figure~\ref{define},
bottom).  This plot was used to guide further spectral extraction (see
\S~\ref{spec_analysis}).  The extended emission along the NW--SE
direction is interesting as this is the direction of the putative
obscuring torus.  Figure~\ref{contour} demonstrates the clear
elongation perpendicular to the bicone (in particular the NW sector),
using the 0.3--2 keV ACIS image smoothed with the CIAO tool $csmooth$.
We extracted the 0.3--2 keV radial profiles of the NW sector, the
bright SW cone, and the faint emission in between (the ``control
region''), shown in Figure~\ref{3profile}.  The NW emission is on
average 5 times fainter than the bright SW cone, but brighter than the
control region at $r<7\arcsec$ ($4\sigma$ significance).  Beyond
$r=7\arcsec$, it becomes indistinguishable with the faint emission.
The implication on the nuclear obscuring structure will be discussed
later in \S~\ref{evidence}.

\begin{figure}
\epsscale{0.8}
%\plotone{resi.ps}
\plotone{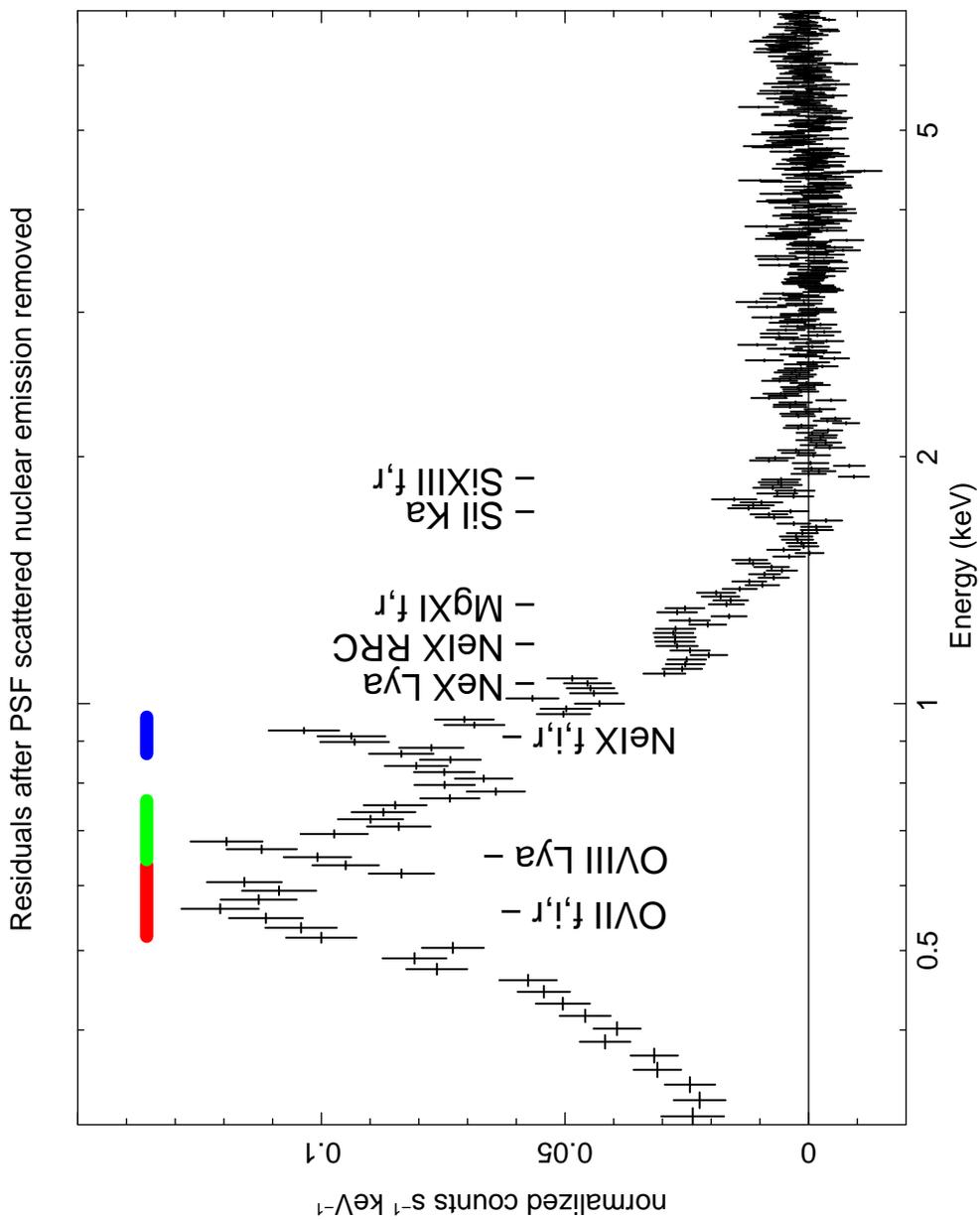}
\caption{The residual spectrum of the large scale extended emission in
  NGC 4151 after the contribution from nuclear emission has been removed,
  demonstrating the presence of X-ray emission lines.  Positions of
  the strong lines from the Ogle et al.\ (2000) HETG spectra are labelled,
  although they are blended at the spectral resolution of ACIS. The
  color bars outline the ranges used for narrow-band line images.
\label{conti}}
\end{figure}

A quick examination of the spectrum of the extended emission extracted
from a circular region between a radius of $2\arcsec$ and $30\arcsec$
shows the clear presence of strong blended emission lines
(Figure~\ref{conti}), corresponding to the emission lines ($<$2~keV)
seen in HETG observations (Ogle et al. 2000).  These lines cannot be
uniquely resolved with the spectral resolution of the ACIS imaging
observation.  Most notably the blended lines appear as three strong
emission features approximately centered at 0.57~keV (OVII), 0.68~keV
(OVIII), 0.91~keV (NeIX) and a few weak lines between 1--2~keV (Mg XI,
Si K$\alpha$, SiXIII; Figure~\ref{conti}).  The blended emission
features seen in the ACIS spectrum and the corresponding HETG lines are
summarized in Table~1 (see \S~2.2.1).

Following Paper II, we created adaptively smoothed images of the NGC
4151 circum-nuclear region using the subpixel resolution data.  We
extracted images from the merged data in three spectral bands
dominated by emission lines below 2 keV: 0.3--0.7 keV (``soft band''
containing the OVII and OVIII emission), 0.7--1.0 keV (``medium band''
containing the NeIX emission), and 1--2 keV (``hard band'').
Figure~\ref{3color}a presents a false color composite image of the
central $\sim 45\arcsec \times 45\arcsec$ ($\sim$3~kpc on a side),
circum-nuclear region of NGC 4151, where the soft, medium, and hard
band smoothed images are shown in red, green, and blue,
respectively. Figure~\ref{3color}b zooms in to the central 1
kpc-radius region, emphasizing the bicone emission.  For visualization
of the faint features, the nuclear region (see Paper II) is saturated
and excluded from this figure. The X-ray biconical morphology has been
described in Paper I, and here we focus on the spectral differences.
The extended bicone emission ($>2\arcsec$) appears rich in 0.3--1 keV
emission (red and green).  Similar color emission (mixed red/green) is
also seen perpendicular to the bicone direction with a hint of harder
(blue) emission.  Clumps of 0.3--0.7 keV (red; OVII and OVIII)
emission are also seen, some particularly prominent features are seen
in the SW cone $\sim 6\arcsec$ and $\sim 10\arcsec$ from the nucleus.

To further highlight the regions where the strong emission lines
arise, we also extracted the X-ray emission in narrow energy intervals
(0.53--0.63 keV; 0.63--0.73 keV; 0.85--0.95 keV) to create line
strength images.  This is a reasonable approach since the line
emission dominates over the weak underlying continuum (Schurch et
al. 2004) in these narrow bands.  The resulting images are shown in
Figure~\ref{lines}.  The position of the nucleus is indicated with a
black cross.  OVII and OVIII line emission is prevalent in the
extended regions. NeIX emission, while extended along the general
P.A. of the large scale extended emission, is bright in the central
$r=2\arcsec$ but becomes fainter quickly at $r\ga 3\arcsec$ compared
to the OVII emission.  We examined similar emission line maps for the
weak lines in the 1--2 keV range (NeX, NeIX RRC, MgXI) and found that
they are highly concentrated within the inner $r=2\arcsec$ region.
This is better demonstrated in Figure~\ref{lines}d, where the radial
profiles of the line emissions are shown.

\subsection{Spectral Analysis}\label{sec-large_scale}

Spectra and instrument responses were generated using CIAO Version 4.2
and analyzed with XSPEC Version 12.6 (Arnaud 1996).  Background
spectra were taken from a source-free region from the same chip.
Spectra were grouped to have at least 20 counts per energy bin to
allow $\chi^2$ fitting. Unless otherwise noted, we restricted our
modeling to photon energies above 0.3 keV, where the ACIS calibration
is good\footnote{See \url{http://cxc.cfa.harvard.edu/cal/Acis/}}, and
below 2 keV, above which the nuclear continuum dominates
\citep{Wang10_NUC}.  The 90\% confidence interval for a single
interesting parameter is reported for all fitting results.

\subsubsection{The $30\arcsec$-radius Extended Emission as a Testbed}\label{brem}

We first characterize the X-ray spectrum of bright extended emission
extracted from the $2\arcsec \leq r\leq 30\arcsec$ circular region,
exploring a phenomenological spectral fitting approach that was
broadly adopted in previous studies at CCD resolution (Yang et
al.\ 2001; see also Ogle et al. 2000, Smith \& Wilson 2001).  The
``continuum'' in the ACIS spectrum (a combination of true underlying
continuum and highly blended weak emission lines) was described by a
smooth bremsstrahlung emission. We emphasize that the bremsstrahlung
emission here should not be assigned any physical origin, whereas the
derived line emission should be considered as a lower limit in such an
approach.

Repeating the technique used to fit the spectrum of the nucleus (Wang
et al.\ 2010), narrow emission lines were then added to the model. The
{\em Chandra} HETG and {\em XMM}-Newton RGS spectra (Ogle et al. 2000;
Schurch et al. 2004; Armentrout et al. 2007) were used as a guide for
the identification of the soft X-ray lines.  The X-ray spectrum of the
nucleus of NGC 4151 is notoriously complex and variable.  The hard
X-ray continuum is dominated by a power law component suppressed by a
partially covering absorber and a Compton reflection component (Weaver
et al. 1994; Schurch et al. 2004; De Rosa et al. 2007; Wang et
al. 2010b; Lubinski et al. 2010), and the absorption consists of both
a mildly-ionized ``warm absorber'' and a cold absorber
\citep{Kraemer05,Omai08}.  The PSF scattered contribution from the
nuclear emission is taken into account in the spectral fitting, using
the nuclear component fixed at the best fit nuclear spectral model and
normalization \citep{Wang10_NUC} scaled to the expected fraction from
PSF simulations.  The extraction region contains 4.7\% of the 0.3--2
keV nuclear emission due to PSF scattering, corresponding to
$\sim$25\% of the extracted counts.  This can be taken as the upper
limit of contamination to the extended line emission, as the line
emission dominates the nuclear spectrum.

The identified emission line blends are summarized in
Table~\ref{tab:extended_flux} and the spectrum shown in
Figure~\ref{4reg2} (see also Figure~\ref{conti}). The absorption
column required by the fit is consistent with the line-of-sight
Galactic column towards NGC 4151 ($N_H=2.1\times 10^{20}$ cm$^{-2}$;
Murphy et al. 1993).  The X-ray luminosity of the extended emission
(between 130 pc and 2 kpc from the nucleus) corrected for the Galactic
absorption is $L_{0.3-2{\rm keV}}=1.1\pm 0.2 \times 10^{40}$ erg
s$^{-1}$, compared with $L_{2-10{\rm keV}}=2\times 10^{42}$ erg
s$^{-1}$ for the unabsorbed nuclear source (Wang et al. 2010b).

However, such a simplified model does not provide constraints on the
emission mechanism, and more physically meaningful spectral models are
needed.  Fitting the data with combinations of absorbed optically-thin
thermal emission with solar abundances ($APEC$ model; Smith et
al.\ 2001) was attempted but gave poor results (reduced $\chi^2_{\nu}
> 3$).  The abundances were then allowed to vary from solar abundance
in the {\tt APEC} model using the variable abundance APEC model ({\tt
  VAPEC}), in which $Z_O$, $Z_{Ne}$, $Z_{Si}$, and $Z_{Fe}$ were left
free while other elements were fixed at solar abundance.  When $Z$ is
allowed to vary, the fit generally improves but still is far from
satisfactory ($\chi^2_{\nu} \sim 2$).  Moreover, the required
abundances become unphysically low ($Z<0.01 Z_{\odot}$) as the fit
attempts to reproduce a smooth continuum.

To have a self-consistent model, we have made use of the {\tt Cloudy}
photoionization modeling code (Ferland et al. 1998).  Although the
resolution of the ACIS CCD spectrum prevents us from performing line-based
diagnostics, e.g. deriving line ratios of the Helium-like triplets,
spectral fitting with photoionization models still provides useful
constraints about the photoionization status (as in Bianchi et
al.\ 2006, 2010; Gonzalez-Martin et al.\ 2010).  Using {\tt Cloudy}
version C08.00, which enables a Cloudy/XSPEC interface (Porter et
al. 2006), we modelled the $<2$~keV X-ray spectrum assuming an open
plane-parallel geometry (``slab'').  The dimensionless ionization
parameter (Osterbrock \& Ferland 2006) is defined as $U=Q/(4\pi
r^2cn_H)$, where $n_H$ is the hydrogen number density, $r$ is the distance
to the inner face of a model slab, $c$ is the speed of light, and
$Q=\int_{13.6eV}^{\infty}{L_{\nu}/h{\nu}}$ is the emitting rate of
hydrogen ionizing photons (s$^{-1}$) by the ionizing source.  We
assumed the broken power-law used by Kraemer et al. (2005) for the AGN
continuum (see also Armentrout et al. 2007) .  We varied $U$ and
$N_H$, the column density of the slab to create spectral models
covering a grid of parameters ($-3 \leq \log U \leq 3$, $19.0 \leq
\log N_H \leq 23.5$ cm$^{-2}$), which were fed to XSPEC.

Armentrout et al.\ (2007) successfully modeled the XMM-Newton/RGS soft
X-ray spectrum of NGC 4151 with three photoionization components: a
high ionization component ($\log U=1.3$, $\log N_H=23.0$), a medium
ionization component ($\log U=0$, $\log N_H=23.0$), and a low
ionization component ($\log U=-0.5$, $\log N_H=20.5$).  In our initial
attempt, we included all these components, with the ionization
parameters and the column densities fixed at these values and the
normalizations set free.  In addition, the photoionized emission was
absorbed with a line-of-sight column density, $N_{H,l.o.s}$.  The fit
to the ACIS spectrum was poor ($\chi^2_{\nu}\geq 6$).  This is not a
total surprise, given that these best fit parameters, are optimized
for the brightest photoionized material close to the nucleus
(10$^{-3}$--1~pc; Armentrout et al.\ 2007) that dominates the soft
X-ray spectrum of NGC 4151, whereas the emission of interest here
arise on the $r\sim$130~pc--1.5~kpc scale.

When the ($U$, $N_H$) parameters for the photoionized components were
set free to vary, we were able to obtain a statistically satisfactory
fit ($\chi^2_{\nu}=1.1$), reproducing the observed ACIS spectrum of
the large scale extended X-ray emission well (Figure~\ref{phot3}).
The line-of-sight column density required by the fit is
$N_{H,l.o.s}=2\times 10^{20}$ cm$^{-2}$, consistent with the Galactic
column towards NGC 4151 \citep{Murphy96}.  Two components, a high
ionization phase ($\log U=0.8$, $\log N_H=20.0$) and a low ionization
phase ($\log U= -0.25$, $\log N_H=19.4$), are required to produce both
the hydrogen-like and helium-like neon and oxygen line emissions.  The
{\tt Cloudy} model predicts gas pressures of $6\times 10^{-7}$ dynes
cm$^{-2}$ and $1.3\times 10^{-7}$ dynes cm$^{-2}$ for the high $U$
component and low $U$ component, respectively.  If they are
collocated, additional pressure--perhaps from a hot intercloud medium
in a multiphase ISM (Elvis et al. 1983, Ogle et al. 2000)--is needed
to reach pressure equilibrium.  Note that the $\log U= -0.25$
component closely resembles the ``Low'' model component in Armentrout
et al.\ (2007) and the ``D+Ea'' X-ray absorption component in Kraemer
et al.\ (2005).

These model components still left small residuals at $\sim$0.5 keV
(the NVII Ly$\alpha$) and 1.02 keV (the NeX Ly$\alpha$).  Adding in a
third photoionized component can further improve the fit (best fit
parameters $\log U= 1.23$, $\log N_H=19.2$), but the improvement
appears not significant when an $F$-test is performed (at $70\%$
significance level).  Thus the two-component model is used as a
template for the following analysis of subregions.  

Given the satisfactory spectral fits with photoionized components and
lack of clear signatures of collisional ionization in the ENLR, how
much thermal emission from hot gas may be present in the
circum-nuclear region of NGC 4151?  We obtain a constraint on the
possible presence of the hot ISM by invoking a collisionally ionized
component in the best fit photoionization model.  After adding a
thermal emission component ($APEC$) to the photoionized emission, the
fit is improved with a $kT=0.81\pm 0.05$~keV (at $>99\%$ significant
level by F-test) and the residual at 0.9--1.1 keV is mostly accounted
for.  This thermal component may contribute $\la 12\%$ of the total
soft X-ray emission, $L_{th,0.5-2{\rm keV}}=6\times 10^{38}$ erg
s$^{-1}$.  As a cautionary note, although $APEC$ component improves
the fit statistically, it remains possible that the residuals are due
to photoexcitation of Fe L-shell and the enhanced resonance lines
(e.g., Ogle et al. 2003; Kinkhabwala et al. 2003) that are
inadequately accounted for with the {\tt Cloudy} photoionization model
components.  This will be explored with the high resolution HETG
spectrum in future work.  We will discuss the possible presence of the
thermal component further.

We emphasize that, because of likely complex geometrical dependencies
and absorption between these components in the kpc scale emission,
this modeling may still be over-simplistic to simultaneously reproduce
the observed features in the broad band.

\subsubsection{Spectral Fitting of Selected Subregions}\label{spec_analysis}

To perform spectral analysis of the extended X-ray emission, the large
scale region was divided into four sectors with position angles (P.A.)
as outlined in Figure~\ref{define}.  Extraction regions are termed the
``SW'', ``NW'', ``SE'' and ``NE'' regions respectively (Table~2; see
Figure~\ref{define}).  The bisectors of NE and SW correspond to the
approximate P.A. of the ENLRs (e.g., Robinson et al. 1994).  They are
further divided into subregions along the radial direction (e.g., SW
1, 2, 3, and 4; NE 1, 2, and 3), with $\sim$1000 counts (0.3--2 keV)
in the fainter outer regions (SW4 and NE3).  The NW and SE sectors
cover regions ($2\arcsec \leq r \leq 15\arcsec$) perpendicular to the
bicone, which is the direction of the large scale weak bar (Asif et
al. 1998).

Corresponding to the variations in color, spectral differences are
clearly visible when the ACIS spectra across sectors are compared
(Figure~\ref{compare4}).  The fitting results are summarized in
Table~2, and Figures~\ref{sw}, ~\ref{ne}, and~\ref{se-nw} show the
corresponding spectra and the fits. We find that:

\begin{enumerate}

\item Except for the outer-most regions (NE3 and SW4), a high
  ionization component is present in all regions ($\log U\sim
  0.83$--1.26).  Lower ionization components are present along the
  bicone direction (NE--SW), with $\log U$ values ranging between -0.5
  and 0.  Recall that the fit to the integrated spectrum, which
  includes emission from all the subregions, finds the $\log U\sim
  0.8$ and $\log U\sim -0.25$ components.

\item In the NE3 and SW4 regions, a very low ionization component
  ($\log U > -1$) and a high column density ($\log N_H=23.5$) are
  required by the fit to improve the residuals in the MgI K$\alpha$
  and SiI K$\alpha$.  An F-test indicates that this fluorescent
  component significantly improve the fit at $98\%$ confidence.

\item The second photoionized component in the NW--SE regions,
  perpendicular to the ionization cone, shows a lower ionization
  ($\log U\sim -1$).

\end{enumerate}

We also extracted spectra of the X-ray emission in the bicone regions
that are prominent in 0.3--0.7 keV emission (or ``OVII'' rich,
visually red in Figure~\ref{3color}) and in 0.7--1 keV emission (or
``NeIX'' rich, visually green in Figure~\ref{3color}). The spectral
fits are shown in Figure~\ref{red-green}a,b respectively.  Whereas the
fitted ionization parameter for the ``NeIX'' rich emission is close to
the average value, the ``OVII'' rich emission is generally of lower
ionization, likely due to higher number density.  This is consistent
with the appearance of clumpy material in the optical.

\section{Discussion}

\subsection{Nature of ENLR Emission: Photoionization and Collisional Ionization}\label{discuss1}

Kiloparsec-scale extended X-ray emission is ubiquitously observed in
nearby obscured Seyfert galaxies with ENLRs (Bianchi et al. 2006;
Guainazzi et al. 2009; Bianchi et al. 2010; Gonzalez-Martin et al.\ 2010),
but the ionization mechanism for this emission has been a subject of
active debate.  For NGC 4151, Yang et al. (2001) proposed a two phase
model for the extended X-ray emission, including a photoionized warm
gas and a hotter collisionally ionized gas (see also Komossa
2001). The relatively low electron temperatures ($T\sim 10^4--10^5$ K)
inferred from the radiative recombination continua (RRC; see Liedahl
1999 for review) and the line ratios from the He-like triplets
strongly support the presence of a photoionized component (Ogle et
al. 2000; Schurch et al. 2004; Kraemer et al. 2005).  However, does
the observed soft spectrum entirely consist of photoionized emission,
or does collisional ionization become more important than the
photoionization mechanism beyond the nuclear region (Armentrout et
al. 2007)?

\subsubsection{Evidences for Photoionization as the Dominant Ionization Mechanism}

Several of our results point to photoionization as the dominant mechanism.

\begin{itemize}

\item{{\em Emission Line Gas Morphologies}} --- First of all, the
  morphological agreement between soft X-ray and optical [OIII] emission is very
  strong. In particular, there is a good spatial correlation between
  [OIII] line emission and the three strong X-ray emission lines
  (Figures~\ref{lines}a-c).  There is clearly an overall morphological
  coincidence with the ionized gas bicone traced by [OIII] emission,
  as noted in previous X-ray studies (e.g., Ogle et al. 2000, Yang et
  al. 2001). A good agreement between the peaks of optical and X-ray
  line emissions in the inner $r=3\arcsec$ is found.

\item{{\em [OIII]-to-Soft X-ray Ratio of the {\em HST}-resolved
    Clouds}} --- Bianchi et al. (2006) studied a sample of eight
  Seyfert 2 galaxies, and suggested that the same kpc-scale gas
  photoionized by the AGN continuum can simultaneously produce the
  observed X-ray and [OIII] emission.  Our HRC and ACIS studies of the
  inner 2\arcsec\/ ($<130$~pc) emission in NGC 4151 (Wang et al. 2009
  and Paper II) have shown that, for most of the clouds close to the
  nucleus, the ratios of [OIII] to soft X-ray emission are consistent
  with the typical ratios in Bianchi et al. (2006), the exception
  being clouds collisionally ionized by the radio jet.

Our subpixel-processed ACIS images of NGC 4151 enable us to directly
compare the extended X-ray emission and those of the resolved [OIII]
clouds (Figure~\ref{lines}).  The main cloud features are labeled in
Figure~\ref{oiii_cloud}.  Using the calibrated continuum-subtracted
HST/WFPC2 F502N image (Kaiser et al.\ 2000), we measured the [OIII]
fluxes for the clouds and listed them in Table~\ref{fluxratio}. ACIS
spectra were extracted from the same regions and fitted with the
continuum plus emission line model to obtain their 0.5--2 keV X-ray
fluxes. Figure~\ref{ratio} shows the [OIII] to soft X-ray ratio for
the distinct cloud features at various radii to the nucleus ($\sim$150
pc--1500 pc).  The ratios listed in Table~\ref{fluxratio} are
consistent with a constant [OIII]/X-ray flux ratio of $\sim$15 for the
1.5 kpc range spanned by these measurements.  The radially constant
[OIII]/X-ray ratios indicate that the relative contributions of the
low- and high-ionization gas phases at different radii are similar, as
pointed out by our referee.  If the [OIII] and X-ray emission arise
from a single photoionized medium, as discussed in Bianchi et
al. (2006), this further implies that the ionization parameter does
not vary significantly to a large radial distance and the density is
approximately decreasing as $r^{-2}$, as expected for a freely
expanding nuclear wind based on mass conservation (Kraemer et
al. 2000; Bianchi et al. 2006).

\item{{\em Spectral Fitting}} --- The spectral analysis shows that, a
  model with just two photoionized components of high- and
  low-ionization parameters can describe well the observed X-ray
  spectra in various regions across the circum-nuclear area, whereas
  in the spectral fits that invoke only collisionally ionized
  components strong emission line residuals persist.  This supports
  the dominant role of nuclear photoionization.

\item{{\em Kinematics of the Ionized Gas}} --- The full velocity field
  of the NLR for NGC 4151 was mapped by Kaiser et al. (2000), which
  supports a hollow biconical outflow model for these clouds.  At
  distances exceeding 4\arcsec\/ ($\sim$260 pc) from the nucleus,
  Kaiser et al. (2000) find that the gas cloud kinematics are
  consistent with circular rotation in a gravitational field,
  identical to the host ISM.  Therefore, except for a few clouds in
  the nuclear region that appear disturbed by the impact of the radio
  jet \citep[][see also Paper II]{Kaiser00,Mundell03}, there is no
  evidence for strong kinematic shocks or any strong shocks in the
  outflow, which is consistent with the gas being photoionized by the
  nuclear radiation.

\end{itemize}

Based on all the above considerations, we conclude that
photoionization by the nucleus is most likely the dominant ionization
mechanism in the ENLR of NGC 4151, confirming the suggestions
of previous studies (Ogle et al.\ 2000; Schurch et al.\ 2004; Kraemer
et al.\ 2005; Armentrout et al.\ 2007; Wang et al.\ 2009b).  It is
reassuring to establish a role for nuclear photoionization in the
extended NLRs that is consistent with the grating studies.  However,
there is also evidence that points to a more complex emission mechanism
than a single photoionized medium (e.g., Bianchi et al.\ 2006). We
discuss these results and their implications below in Sections~3.2
through~3.4.

\subsubsection{Constraints on the Presence of Collisionally Ionized Gas}

In \S~\ref{sec-large_scale}, we estimated the extent of thermal
emission from hot gas ($L \la 6\times 10^{38}$ erg s$^{-1}$) that
could be present in the circum-nuclear region of NGC 4151.  The {\em
  Chandra} grating study of NGC 4151 by Ogle et al.\ (2000) provided
the first direct evidence of X-ray line emission from a collisionally
ionized hot plasma.  Armentrout et al.\ (2007) also noted that the
poor fit to the OVII line profile could be due to a non-photoionized
component.  Wang et al. (2011b) further show that a $kT\sim 0.6$ keV
thermal emission component is present in the inner $r=2\arcsec$
region, which likely originates from jet-cloud interaction.

Although star formation activity in the circum-nuclear region is at a
low level (Asif et al.\ 2005; Dumas et al. 2010), hot gas may
originate from the NGC 4151 galactic disk, since its temperature
($kT\sim 0.8$~keV) and luminosity ($L \sim 6\times 10^{38}$) are
comparable to the disk diffuse emission in some star forming galaxies,
$kT\sim 0.6$~keV, $L\sim 10^{39}$ erg s$^{-1}$ (e.g., NGC 253;
Strickland et al.\ 2002).

Another plausible alternative is that the hot plasma is the intercloud
medium uniformly filling in the outflow (i.e., filling factor $f=1$),
providing pressure confinement for the NLR clouds (Elvis et al. 1983;
cf. Krongold et al. 2007 for warm absorber outflows).  The volume
emission measure for the thermal component is $n_e^2 V=4\times
10^{61}$ cm$^{-3}$, derived from the normalization after fitting an
additional thermal component (APEC) to the photoionization models of
the bright biconical emission (\S~2.2.1).  Assuming the hollow
bi-conical NLR geometry (Crenshaw et al. 2000; Das et al. 2005), we
obtain $V\sim 10^{61}$ cm$^{-3}$ and so an average density of $\sim 2$
cm$^{-3}$.  Thus the thermal pressure of the $T\sim 10^7$ K gas is
$\sim 5.5\times 10^{-9}$ dyne cm$^{-2}$.  Comparing with the thermal
pressure of $T\sim 10^4$~K and $n=1600$ cm$^{-3}$ NLR clouds (Robinson
et al.\ 1994) gives $p=4.8\times 10^{-9}$ dyne cm$^{-2}$. Thus
pressure equilibrium between the hot and cool phase is highly likely,
confirming the findings on pressure confinement in Ogle et
al. (2000).

\subsection{A Complex Photoionized Medium: Displacement of Optical and X-ray Features}\label{evidence}

Figure~\ref{lines}a clearly shows that one distinct OVII emission blob
``A'' (at $6\arcsec$, $\sim$390 pc SW from the nucleus) does not align
with a bright [OIII] emitting cloud, designated as ``K1'' in Robinson
et al.\ (1994) and offset by $1\arcsec$ to the SE of ``A'' (see Yang
et al. 2001).  We find this is also the case for a more distant X-ray
blob ``B'' ($10\arcsec$ SW) where the closest [OIII] emission cloud is
2\arcsec\/ to the SW.  Similar displacements between the X-ray
enhancement and the [OIII] blob are also clearly found in the OVIII
line emission (Figure~\ref{lines}b).  Note that the {\em Chandra}
astrometry is accurate to 0.3\arcsec (Paper I), therefore the
mis-alignment between features at this spatial scale is significant
and indicates that the origin of the optical/X-ray emission needs
further investigation.

The radial location of X-ray blob``A'' lies at the boundary between
NLR and ENLR identified in the optical (Robinson et al.\ 1994), where
the density drops rapidly from $n\sim 1600$ cm$^{-3}$ to $n\sim 200$
cm$^{-3}$ \citep{Penston90,Robinson94}.  We can estimate the
ionization parameter of K1.  From the nuclear spectrum
\citep{Wang10_NUC}, we obtain an ionizing luminosity
$L_{13.6eV-100keV}\sim 10^{44}$ erg s$^{-1}$.  Assuming a number
density of 200 cm$^{-3}$ (Penston et al.\ 1990; Robinson et al.\ 1994)
and a radial distance of $1.2\times 10^{21}$ cm to the nucleus, we
obtain an ionization parameter $\log U=-2$.  This is lower than, but
in agreement with, the value of $\log U\sim -1.7$ derived from the
[OIII]/[OII] ratio in Robinson et al.\ (1994), supporting the
photoionization assumption.

Since the X-ray blob and the [OIII] cloud are located at the same
radial distance, their spatial offset suggests that the X-ray blob is
in a higher ionization phase than the [OIII]-bright cloud.  A likely
explanation is that the X-ray blob has a lower density in the clump.
If their densities are similar, the difference might arise from a higher
ionizing flux received at the surface of the X-ray blob, perhaps due
to less screening than in the direction towards the [OIII] cloud.

\subsection{The Leaky ``Torus''}\label{torus}

Our spectral fitting results show that a range of ionization states
are needed to obtain good fits, in agreement with the multiple
ionization parameters required in the grating spectra (e.g.,
Armentrout et al.\ 2007).

A high ionization ($\log U \sim 1$) component is found in the spectral
fitting for all regions surrounding the nucleus, including the NW and
SE sectors perpendicular to the ionization cone (Table~2).  A low
ionization ($\log U \sim -1$) component is also present in the NW--SE
direction that has lower $U$ than found in the cone direction
(NE--SW).  This strongly indicates that there is no continuous
absorbing ``torus'' blocking all the ionizing photons. This hypothesis
is supported by the agreement between the observed H$_2$ line flux in
the ``torus'' direction and the value predicted from X-ray-irradiation
by the active nucleus (Paper I), and by the measurement of HI
absorption in the nuclear region, where high levels of clumping on the
smallest scales is found (Mundell et al.\ 2003). We note that a
``patchy torus'' or clumpy wind arising from the accretion disk in
which the clouds are optically thick \citep[e.g.,][]{Elitzur06}, may
allow leakage of nuclear continuum for such a high $U$ component.
However, it cannot reproduce the $\log U \sim -1$ component in the
directions where the continuum is blocked.

Extended emission along the ``torus'' direction is clearly present in
the large scale X-ray images shown in Figures~\ref{contour} (indicated
by the arrows) and ~\ref{3color}a.  The base of this emission is at
the location of the nucleus, thus may indicate ionization by the
leaked photons in this direction.  Moreover, we found an intriguing
coincidence between radial regions of enhanced X-ray surface
brightness and the P.A. of outflowing optical clouds identified by the
{\em HST} study (Das et al.\ 2005).  They pointed out that these faint
clouds flowing close to the plane of the torus would need to be
ionized by leaked nuclear radiation, which is attested by our X-ray
data.  The low $U$ suggests a lower ionizing flux received by some
clouds, likely due to nuclear continuum filtered by ``warm absorbers''
\citep[e.g.,][]{Krongold07} or the cone walls
\citep[e.g.,][]{Kraemer08}.  Evaluating the difference between the
ionization parameters in the direction of torus and in the bicone, we
estimate that approximately 1\% of the total ionizing flux is seen by the
NW--SE clouds.

\subsection{AGN Interaction with the Host Galaxy Disk}

Previous work on the NLR kinematics of NGC 4151 have established that
the bi-cone geometry in NGC 4151 results in one side of the conical
outflow exiting the galactic plane at a small angle, partly
intersecting the host disk (e.g., Robinson et al.\ 1994; Crenshaw et
al.\ 2000; Das et al.\ 2005; Storchi-Bergmann et al.\ 2009).  Thus
part of the host ISM is likely illuminated by the AGN radiation.  Our
X-ray work finds that multiple phases of photoionized gas are present
in the ENLR of NGC 4151.  The fitting results for the ``OVII''
rich (red in Figure~\ref{3color}) and ``NeIX'' rich (green in
Figure~\ref{3color}) regions indicate that there is a range of ionizations,
with a lower ionization phase associated with the clumpy clouds and
the higher ionization with the diffuse outflow.  There is a lowest
ionization ($\log U \sim -2$) component at larger radii in the SW and
NE sectors.  This component may come from high density material in the
host disk plane, perhaps the nuclear spirals, illuminated by the
nuclear radiation.

We further note that interactions between the biconical outflow and
the host galaxy ISM, similar to the radio jet--cloud interaction in
the nuclear region, may contribute to the heating of any hot gas
present in the host disk.  Currently there is no evidence in the gas
kinematics that supports any strong interaction (Kaiser et al.\ 2000).
However, as we discuss below, there is some morphological evidence
that, at the largest radii, the outflow runs into a clump in the
inhomogeneous ISM and perhaps results in local shock heating of the
gas.

Figure~\ref{morph} gives a large scale (1~arcmin$\times$1~arcmin) view
of the key ISM components in the circum-nuclear region of NGC 4151,
illustrating the directions of the kinematic major axis of the
host galaxy (green solid line; P.A.$\sim$22$^{\circ}$, Pedlar et
al. 1992; Mundell et al. 1999), the large scale ``weak fat bar''
(green dotted line; P.A.$\sim$130$^{\circ}$; Mundell \& Shone 1999),
the ENLR bicone (cyan line; P.A.$\sim$65$^{\circ}$, Evans et
al. 1993), and the distribution of HI (Mundell et al. 1999) and CO
(Dumas et al. 2010) gas, respectively.

In Paper I we noted that the NE soft X-ray emission reaches as far as
12\arcsec\/ (780 pc) from the nucleus, but that the surface brightness
then decreases significantly by a factor of 5.  Comparing the
continuum-subtracted H$\alpha$ emission (Knapen et al. 2004) with the
extended soft (0.3--1 keV) X-ray emission (Figure~\ref{fig8}a), the
ionized gas traced by bright H$\alpha$ emission is mainly located in
the $r\sim 10\arcsec\/$ biconical region along the NE--SW direction
centered on the nucleus, which also closely follows the soft X-ray
emission.  An arc-like feature in the H$\alpha$ image ($\sim 15\arcsec
\times 2\arcsec$) is clearly present at the $r=10\arcsec$ location
(Figure~\ref{morph}) at the edge of the X-ray emission, and the
northern CO gas lane is also found here (Figure~\ref{fig8}b, and Dumas
et al. 2010).

Figure~\ref{fig8}b provides a multiwavelength, enlarged view of this
region, showing the spatial relations between the soft X-ray emission,
the H$\alpha$, the CO gas lanes, and the HI distribution \citep{MS99}.
The termination of the NE X-ray emission appears to be closely
associated with the presence of the dense cold neutral material
(traced by the CO and HI emission) in the galactic disk.  No similar
feature is seen to the SW of the nucleus.

The origin of the H$\alpha$ arc (Figure~\ref{fig8}a) at the terminus
of the NE X-ray emission is intriguing.  It is spatially very close to
the northern CO gas lane and is one of the dust arcs identified in the
optical $V-I$ color map \citep{Asif98}, which are gaseous compressions
driven by the bar potential \citep{Mundell99,Dumas10}.  In this
context, the H$\alpha$ arc may be a more clumpy continuation of the
dense CO arc and photoionized by the AGN radiation.  The rate of
ionizing photons from the AGN is approximately $10^{53}$ s$^{-1}$
(Kraemer et al.\ 2005; c.f. $L_{13.6eV-100keV}\sim 10^{44}$ erg
s$^{-1}$, Wang et al. 2010b).  Assuming that the arc has a
line-of-sight depth that is comparable to its projected length
($\sim$1 kpc), the arc's intercepted ionizing photon rate is $8\times
10^{51}$ s$^{-1}$ at its distance to the nucleus (970 pc), a covering
factor of almost $10\%$.  This photon budget should be treated as an
upper limit as some fraction of the ionization photons will be
absorbed or scattered on the way to the arc.  If we assume that the
gas is in photoionization equilibrium, for the case B effective
recombination rate of H$\alpha$ \citep{Oster89}, a rate of $8\times
10^{49}$ s$^{-1}$ ionizing photons s$^{-1}$ is needed to produce the
observed H$\alpha$ luminosity of the arc ($L_{\rm{H}\alpha} \sim
8\times 10^{37}$ erg s$^{-1}$), which is only 1\% of the maximal rate
of available ionizing radiation in the direction of the arc (for a
covering factor $f_c=10\%$).  The scenario where part of the CO gas
lane is photoionized by the AGN and produces the H$\alpha$ arc seems
plausible with a modest covering factor of $f_c \sim 0.1\%$, implying
that the arc has a filamentary structure with a thickness of a few pc.

On the other hand, we consider the possibility that the arc could be a
bow shock feature from interaction between the biconical outflow and
dense gas in the host disk piled in the CO gas lanes by the large
scale stellar bar.  Assuming that the X-ray emission is due to shock
heating as the outflow encounters the dense CO lane, and that the
shock is strong and adiabatic, we estimate the shock velocity $v_s$,
adopting a postshock temperature $T_{ps}=(3/16)(\mu v_s^2/k_B)$
(Lehnert et al.\ 1999), where $\mu$ is the mass per particle (CO and
HI) and $k_B$ the Boltzmann constant.  For the $kT\sim 0.3$~keV X-ray
temperature measured at the arc, the required $v_s$ is $\sim$150 km
s$^{-1}$, which is much higher than the local sound speed $c_s=10$~km
s$^{-1}$ in the $\sim 10^4$~K H$\alpha$-emitting gas.  Although there
is no direct measurement at the exact position of the H$\alpha$ arc,
the measured outflow velocity in [OIII] at $r=8\arcsec$ NE of the
nucleus is fully consistent with this value of $v_s$, e.g., cloud\#28
which has $v=190\pm 32$ km s$^{-1}$ (Kaiser et al. 2000).

\subsection{Mass Outflow and Kinematic Power}

For two of the brighter, spatially resolved [OIII]-emitting clouds
seen in the $HST$ images (clouds \#1 and \#2 in
Figure~\ref{oiii_cloud}; see also Figure 4 in Kaiser et al. 2000), we
were able to constrain the number density for the X-ray emitting gas
with {\tt Cloudy} modeling of their X-ray spectra when an inner radius
$r=100$ pc to the nucleus and an ionizing photon rate $Q=10^{53}$ are
set.  The best-fit density is $\log n_H=2.9\pm 0.5$ cm$^{-3}$.  We use
this number density to estimate the mass loss rate and the kinematic
power of the hot phase outflow, which are key quantities that gauge
the impact of AGN outflow on the host galaxy ISM.

Following Barbosa et al. (2009) and \citet{SB10}, we calculate the
mass loss rate using $\dot{M}_{x}$ = $ n_H m_H v_r C_g A$, where $v_r$
is the radial velocity (taken as the approximate outflow velocity),
$m_H$ the proton mass, $C_g$ the gas filling factor, and $A$ the area
in the cross section of the outflow.  For a radial distance of $r =
4\times 10^{20}$ cm (130 pc) and the hollow cone geometry of Das et
al.\ (2005), we obtain $A= \pi r^2 (sin^2\theta_{out}-
sin^2\theta_{in})= 1.2\times 10^{40}$ cm$^2$.  Adopting $C_g=0.11$
\citep{SB10} and $v_r \sim 750$ km s$^{-1}$ (Crenshaw et al. 2000;
Storchi-Bergmann et al. 2010), we obtain a mass outflow rate
$\dot{M}_{x}=2.1 M_{\odot}$ yr$^{-1}$ and a kinetic luminosity for the
outflow $L_{outflow}=1/2\dot{M}_{x}v^2 \sim 1.7 \times 10^{41}$ ergs
s$^{-1}$, or $\sim 0.3\%$ of the bolometric luminosity of the AGN
($L_{bol}=7.3\times 10^{43}$erg s$^{-1}$; Kaspi et al. 2005).

The $\dot{M}_{x}$ derived from the hot phase outflow is comparable to
the value derived from NIR studies of ionized gas in NGC 4151
\citep{SB10}, which estimated mass outflow rate $\approx$1.2
$M_{\odot}$ yr$^{-1}$ along each cone.  The $\dot{M}_{x}$ value is
over 10 times higher than the outflow rate derived from $UV$ and
optical spectra (0.16 $M_{\odot}$ yr$^{-1}$; Crenshaw \& Kraemer
2007), but the latter is measured much closer to the nucleus (at 0.1
pc).  The optical emitting gas likely has a much smaller filling
factor than the X-ray emitting gas, which explains the difference in
the measurements.  If our study measured an outer part of the same
outflow, it could have accelerated to a higher velocity or have loaded
ISM between 0.1 pc and 100 pc, hence a higher $\dot{M}_{x}$.

How does $\dot{M}_{x}$ compare to the accretion rate of the SMBH in
NGC 4151? The bolometric luminosity of NGC 4151, $L_{bol}=7.3\times
10^{43}$ erg s$^{-1}$ (Kaspi et al. 2005), is about 1.2\% of its
Eddington luminosity, which is $L_{Edd}\sim 6\times 10^{45}$ erg
s$^{-1}$ for the black hole mass $M_{BH}=4.57^{+0.57}_{-0.47}\times
10^7$M$_{\odot}$ of NGC 4151 (Bentz et al.\ 2006).  Adopting
$L_{bol}=\eta \dot{M}_{accr} c^2$ and $\eta=0.1$ for the efficiency of
the accretion disk (Shakura \& Sunyaev 1973), this implies that the
central engine of NGC 4151 is accreting at $\dot{M}_{accr}=0.013
M_{\odot}$ yr$^{-1}$.  Since the measured outflow rate is 160 times
the accretion rate necessary to feed the active nucleus, there must
be either significant entrainment of the host galaxy's ISM while the
nuclear outflow is expanding, or part of the gas fueling the nucleus
is ejected before radiating, or a high fraction (over 99\%) of the
X-ray gas does not take part in the outflow and is illuminated by
the AGN ``in situ''.  However, the latter seems unlikely as the
grating spectra strongly indicate that the emission lines have outflow
velocities of 200--300 km s$^{-1}$ (e.g., Armentrout et al.\ 2007).
Significant mass loading of the outflow by the host ISM may be most
plausible and common, since previous estimated NLR outflow rates in
Seyferts ($\sim$0.1--10 $M_{\odot}$ yr$^{-1}$) generally exceed
tenfold the accretion rate (Veilleux et al.\ 2005).

Nevertheless, even with such a high $\dot{M}_{x}$ in NGC 4151, the
kinematic power $L_{outflow}$ suggests that only 0.3\% of the
available accretion power is extracted to drive the outflow, similar
to the findings of \citet{SB10} and Holt et al.\ (2006, 2011).  For a
sample of Seyfert NLR outflows (NGC 4151 not included) studied in the
[SIII]$\lambda 9069$ emission, Barbosa et al. (2009) reported a lower
$L_{outflow}/L_{bol}=10^{-5}-10^{-4}$.  As pointed out in Mathur et
al.\ (2009), this poses a problem to the majority of quasar feedback
models which require a higher fraction of the accretion power of the
black hole ($\sim$5\% of $L_{bol}$ to unity) to be thermally coupled
to the host ISM.  Our $L_{outflow}/L_{bol}$ is in rough agreement with
a two-stage feedback model recently proposed by Hopkins \& Elvis
(2010), which requires only 0.5\% of the radiated energy to drive the
initial hot outflow to efficiently shut down star formation in the
host.

\section{Conclusions}

In this paper we present spectral analysis and emission line images
from deep {\em Chandra} observation of NGC 4151, aiming to resolve and
characterise the X-ray emission in the NLR.  The findings are
summarized as follows:

\begin{itemize}

\item The soft X-ray emission line images of NGC 4151 (OVII, OVIII,
  and NeIX) are clearly extended and show remarkable morphological
  coincidence with the biconical NLR mapped by the [OIII] emission,
  which supports a common emission mechanism for the hot- and
  cool-phase of the NLR gas.

\item Extended emission in the X-ray image is detected along the
  NW--SE sectors, which is the direction of a putative torus. This may
  explain the faint rogue clouds identified in previous HST studies
  that require ionization in this direction, indicating leakage of
  nuclear ionization instead of full blocking by a continuous
  obscuring torus.

\item Spectral models involving smooth continua (a bremsstrahlung
  plus a power law) with emission lines provide good descriptions of
  the spectra.  The emission lines cannot be uniquely identified with
  the present spectral resolution, but are consistent with the
  brighter lines seen in the {\em Chandra} HETGS and XMM-Newton RGS
  spectra below 2 keV.  The absorption corrected X-ray luminosity of
  the extended emission between $r=130$ pc and $r=2$ kpc is
  $L_{0.3-2{\rm keV}}=1.1\pm 0.2 \times 10^{40}$ erg s$^{-1}$.

\item Photoionization models successfully reproduce the soft X-ray
  emission, supporting the dominant role of nuclear photoionization.
  There are considerable variations in ionization states across the
  circum-nuclear region.  A high ionization ($\log U\sim 1$) component
  is present in most regions.  A low ionization ($\log U\sim -0.25$)
  component is present along the bicone direction (NE--SW), whereas a
  lower ionization ($\log U \sim -1$) component is found in the NW--SE
  direction, which is consistent with filtered nuclear emission by
  warm absorbers instead of a continuous absorbing torus.  The lowest
  ionization component ($\log U \sim -2$) appears to be associated with the
  dense gas in the host plane.  A thermal component may still be
  present at $\la 12\%$ of the total soft emission, perhaps related to
  hot ISM in the galactic disk or shocks associated with the outflow.

\item The measured ratios of [OIII]/soft X-ray flux are consistent
  with a constant ratio of $\sim$15 for the 1.5~kpc radius spanned by
  these measurements.  This suggests a similar relative contributions
  from the low- and high-ionization gas phases at different radii.  If
  the [OIII] and X-ray emission arise from a single photoionized
  medium, this further implies an outflow with a wind-like density
  profile ($n_e\propto r^{-2}$).

\item The estimated mass outflow rate in NGC 4151 is
  $\sim$2$M_{\odot}$ yr$^{-1}$ at 130 pc and the kinematic power of
  the ionized outflow is $1.7\times 10^{41}$ erg s$^{-1}$, 0.3\%
  $L_{bol}$.  This value is significantly lower than the expected
  efficiency in the majority of quasar feedback models, but comparable to
  the two-stage model described in \citet{Hopkins10}.

\end{itemize}

Placing all our findings in the context of previous studies, we obtain
a comprehensive view of the NGC 4151 circum-nuclear region at various
spatial scales, from the inner-most $r\sim 50$ pc (as illustrated in
the schematic drawing in Figure~\ref{cartoon}a) to as far as $r\sim 2$
kpc (Figure~\ref{cartoon}b).  The key points are recapped here.

Photoionization by the nucleus is important at all scales. In the
nuclear region, except for a few clouds that show interaction with the
radio jet (Wang et al. 2009, 2011b), most of the X-ray NLR clouds are
consistent with being part of a photoionized biconical outflow, with a
$n\propto r^{-2}$ density profile expected for a nuclear wind.

In the ENLR of NGC 4151, photoionization is still the dominant
ionization mechanism for the observed X-ray emission (e.g., Ogle et
al. 2000; Kraemer et al. 2005), but a wide range of ionization states
is present.  This is likely related to the nature of the ENLR.  The
bi-cone geometry indicates that one side of the conical outflow exits
the galactic plane at a small angle, partly intersecting the host disk
(e.g., Crenshaw et al.\ 2000; Das et al.\ 2005; Storchi-Bergmann et
al. 2010).  Optical studies (e.g., Robinson et al. 1994; Kaiser et
al. 2000) have identified such a boundary between NLR and ENLR at
$r\approx 6\arcsec$ (SW), and the ionized gas kinematics become
consistent with the rotation velocity of the galactic disk at $r\geq
6\arcsec$.  Thus the origin of the ENLR is best explained as the
inhomogeneous ISM in the galactic plane ionized by the AGN.  Part of
the high density nuclear spirals illuminated in the path of the bicone
can produce the curvy shape of X-ray enhancement and the associated
low ionization.  The presence of ionized emission perpendicular to the
bicone indicates leaked nuclear emission along the putative torus
direction, likely filtered nuclear emission by warm absorbers.

In the large 3~kpc-scale cavity of the H I material, we find faint
soft diffuse X-ray emission that provides evidence for AGN--host
interaction in NGC 4151 (Wang et al. 2010a), originated from either
hot gas heated by the nuclear outflow or photoionized gas from a
recent nuclear outburst.

As a concluding remark, our findings in NGC 4151 demonstrate that
abundant information can be extracted from X-ray spectral imaging
studies when supplemented with multiwavelength data.  Many valuable
new radio, optical and IR observations and modeling of other nearby
Seyfert galaxies are emerging (Crenshaw et al. 2010; Riffel et
al. 2010; Fisher et al.\ 2011, Schnorr M{\"u}ller et al.\ 2011), and
progress in the X-rays are being made to further our understanding of
AGN feeding and feedback processes in these galaxies.

\acknowledgments

We thank the anonymous referees for providing us with detailed and
constructive comments that have improved the clarity of this
manuscript.  This work is supported by NASA grant GO8-9101X and
GO1-12009X.  We acknowledge support from the CXC, which is operated by
the Smithsonian Astrophysical Observatory (SAO) for and on behalf of
NASA under Contract NAS8-03060.  CGM acknowledges financial support
from the Royal Society and Research Councils U.K. GD was supported by
DFG grants SCH 536/4-1 and SCH 536/4-2 as part of SPP 1177.
J. W. thanks T. Kallman, G. Ferland, S. Bianchi, A. Marinucci, and
S. Chakravorty for advice on photoionization modeling.  This research
has made use of data obtained from the {\em Chandra} Data Archive, and
software provided by the CXC in the application packages CIAO and
Sherpa.  Some of the data presented in this paper were obtained from
the Multimission Archive at the Space Telescope Science Institute
(MAST). STScI is operated by the Association of Universities for
Research in Astronomy, Inc., under NASA contract NAS5-26555.

{\it Facilities:} \facility{CXO (HRC, ACIS)}

\clearpage

\clearpage

\clearpage

\begin{figure}
\epsscale{0.7}
%\plotone{3color_faint.ps}
\plotone{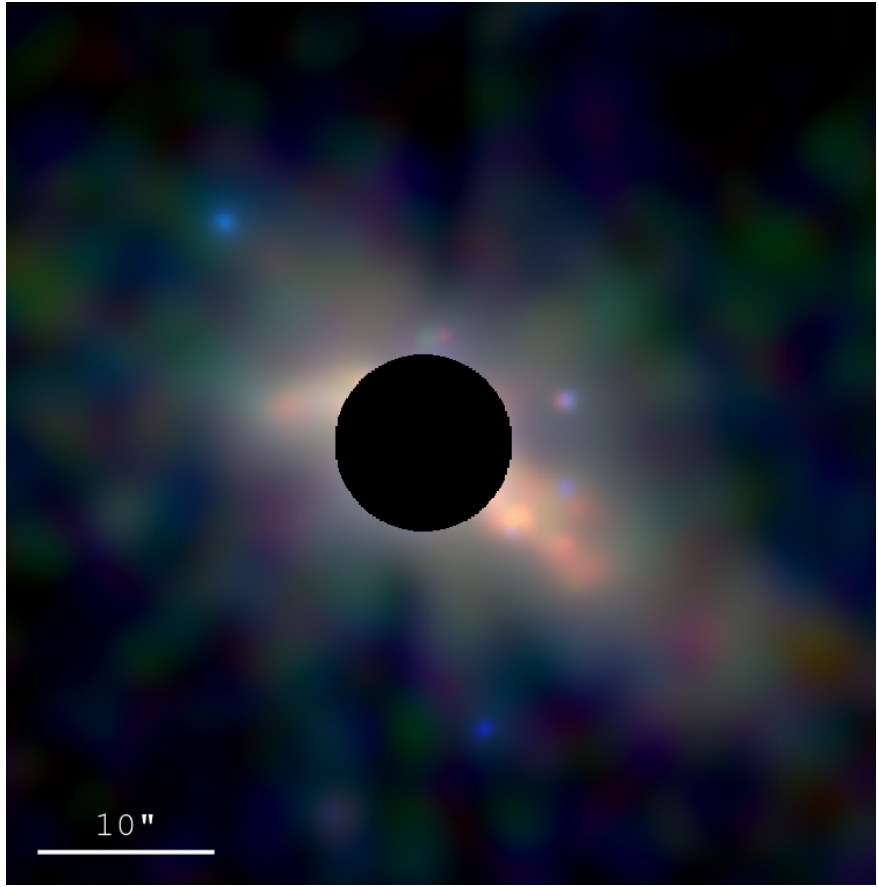}
\caption{(a) Tri-color composite image of the central 3 kpc-diameter
  region of NGC 4151, where the soft (0.3--0.7 keV), medium (0.7--1
  keV), and hard band (1--2 keV) adaptively smoothed images are shown
  in red, green, and blue, respectively. The inner $r=5\arcsec$ region is
  intentionally saturated and masked to show the faint features
  farther out. (b) Same as Figure~\ref{3color}, but for the inner 1 kpc-radius
  region of NGC 4151.  The inner $r=2\arcsec$ nuclear region is
  masked.
\label{3color}}
\end{figure}

\addtocounter{figure}{-1} 
\begin{figure}
\epsscale{0.7}
\plotone{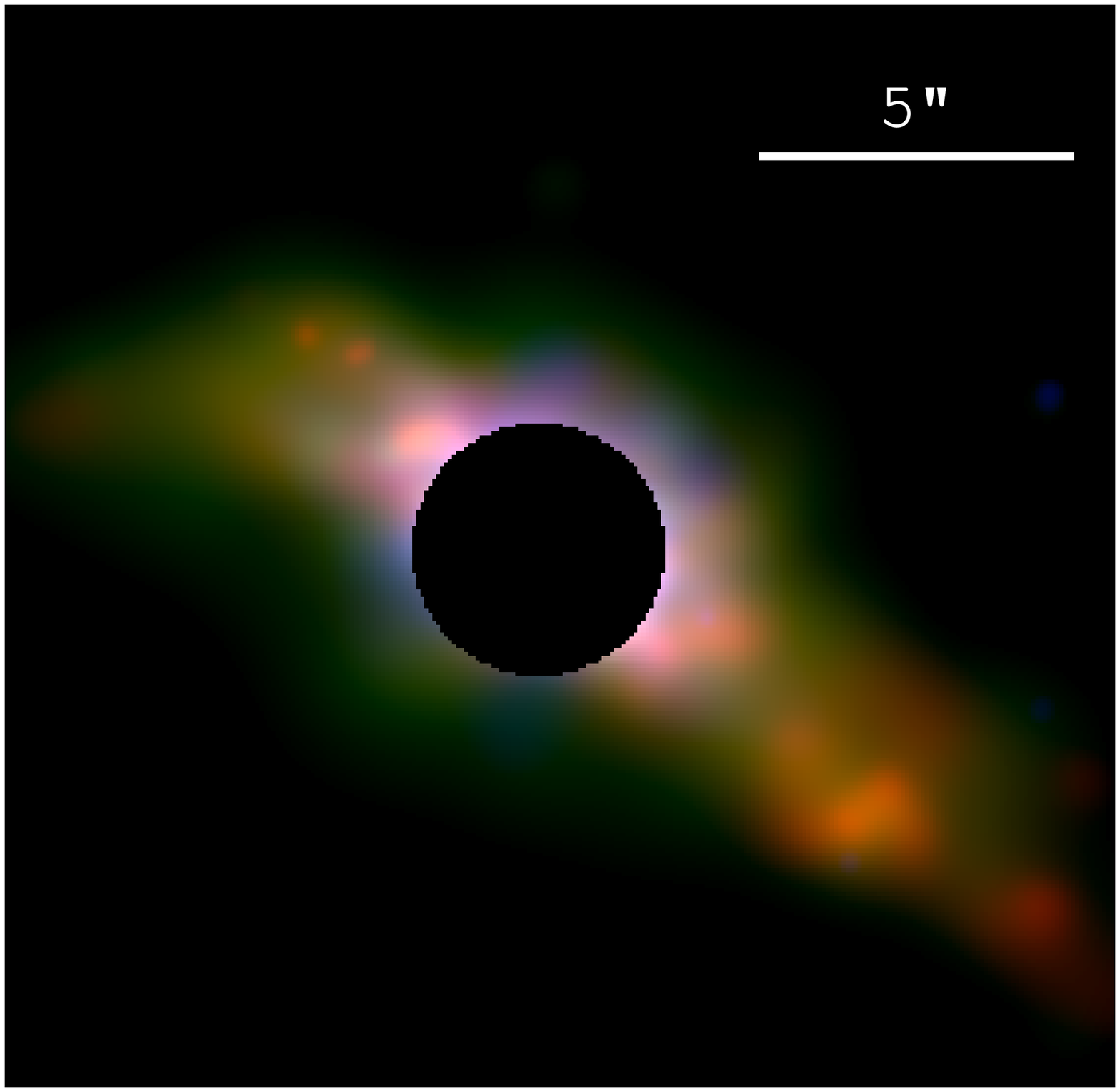}
\caption{---Continued.}
\end{figure}

\begin{figure}
\epsscale{0.6}
\vspace{-0.3in}
%\plotone{O3_lines2.ps}
\plotone{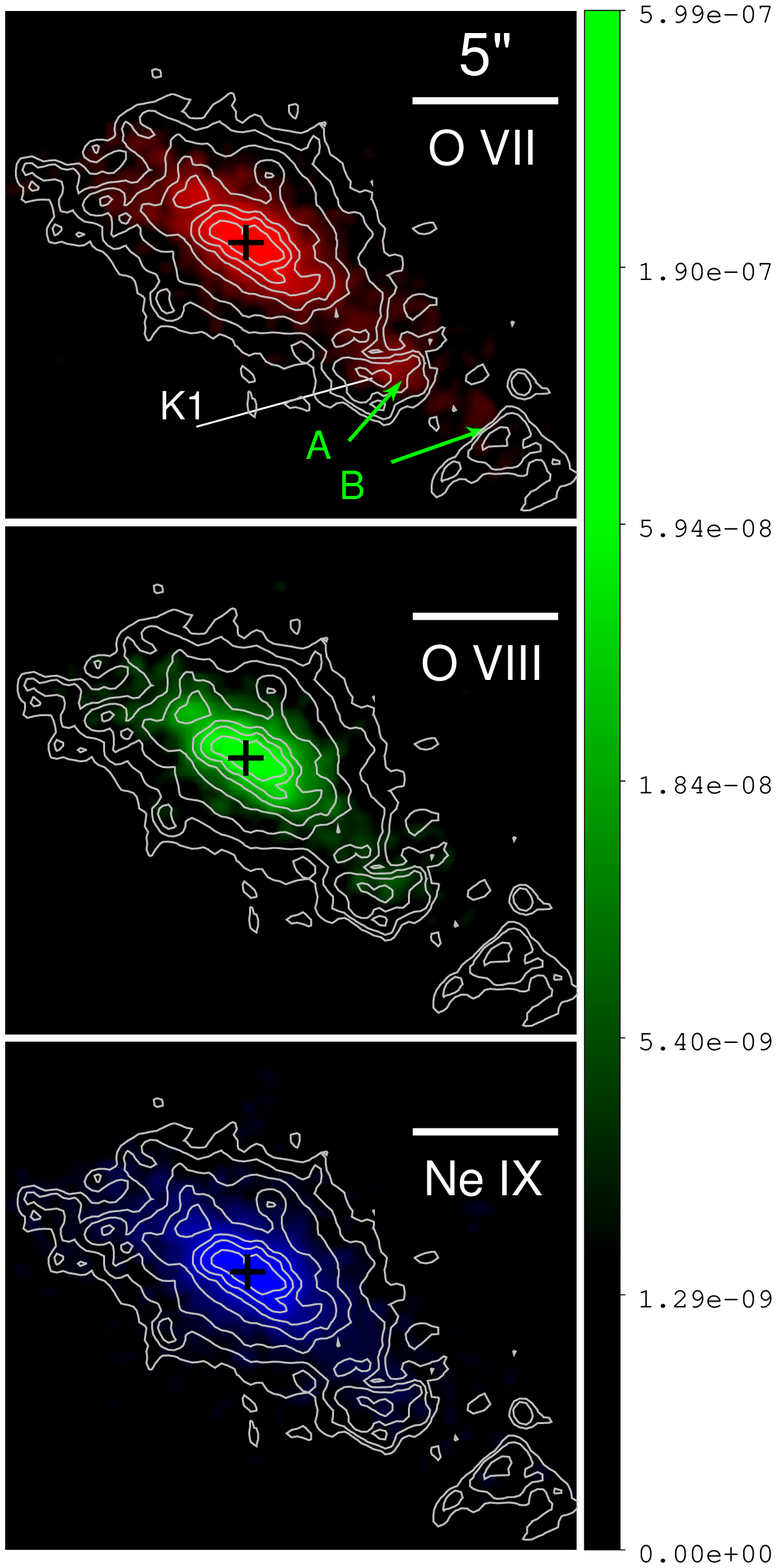}
\vspace{-0.3in}
\caption{The narrow-band ACIS images of the three strongest emission
  features in the central 15\arcsec\/ ($\sim$1 kpc-across) region of
  NGC 4151 (a) OVII; (b) OVIII Ly$\alpha$+OVII RRC; (c) NeIX OVII
  triplet line emission.  Overlaid are contours of the {\em HST} WFPC2
  [OIII] $\lambda$5007\AA\/ image (Kaiser et al. 2000). The position
  of nucleus is indicated with a cross. ``A'' and ``B'' mark the X-ray
  structures displaced from optical clouds.  ``K1'' is a bright [OIII]
  cloud identified in Robinson et al.\ (1994). See
  \S~\ref{evidence}. (d) The radial profile comparison for the
  emission line fluxes with 1-$\sigma$ error bars, including the
  weaker emission lines NeX and Mg XI (see the legend in the top-left
  panel).
\label{lines}}
\end{figure}

\addtocounter{figure}{-1} 
\begin{figure}
\epsscale{0.7}
\plotone{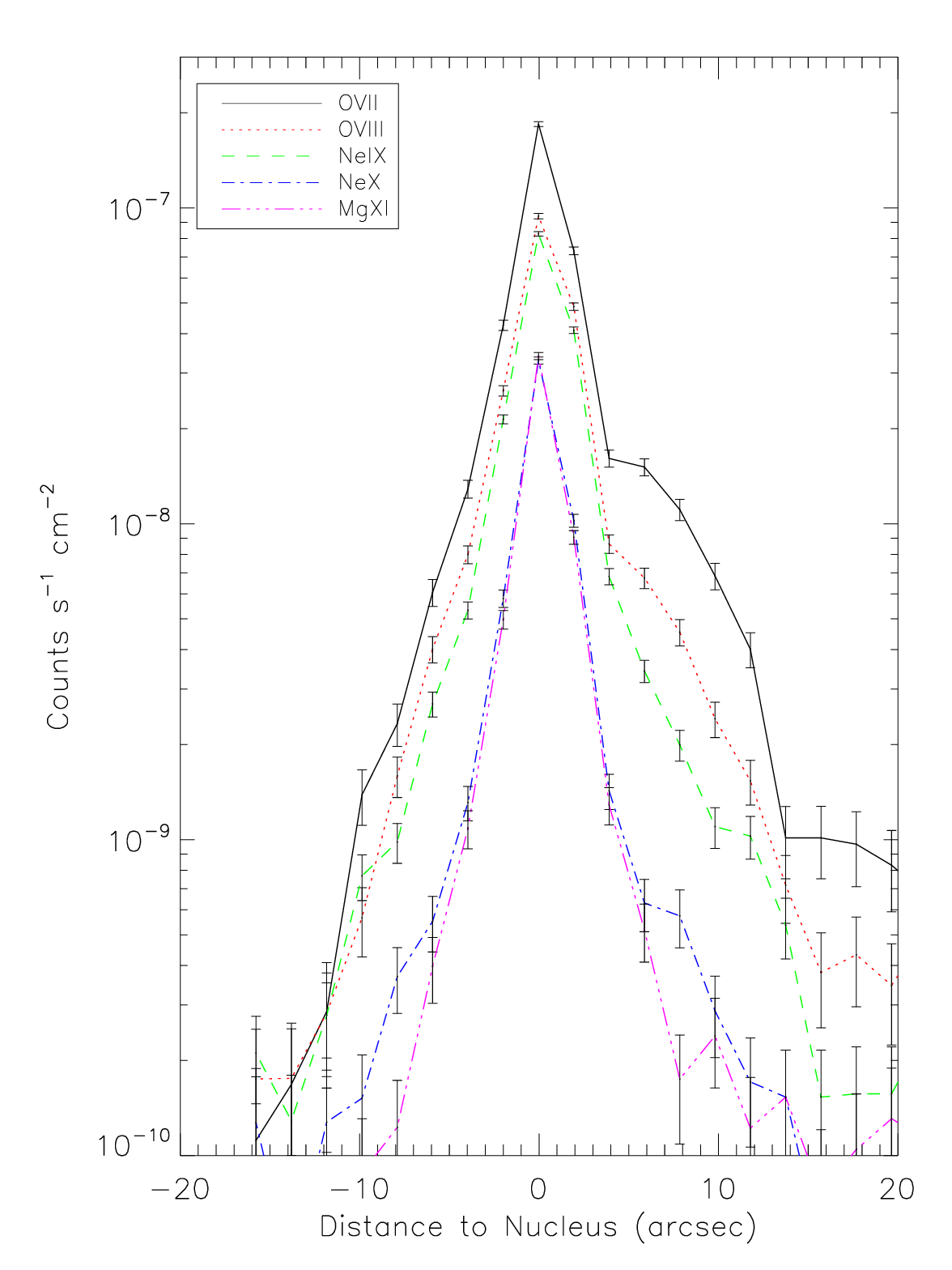}
\caption{---Continued.}
\end{figure}

\begin{figure}
\includegraphics[scale=.5,angle=-90]{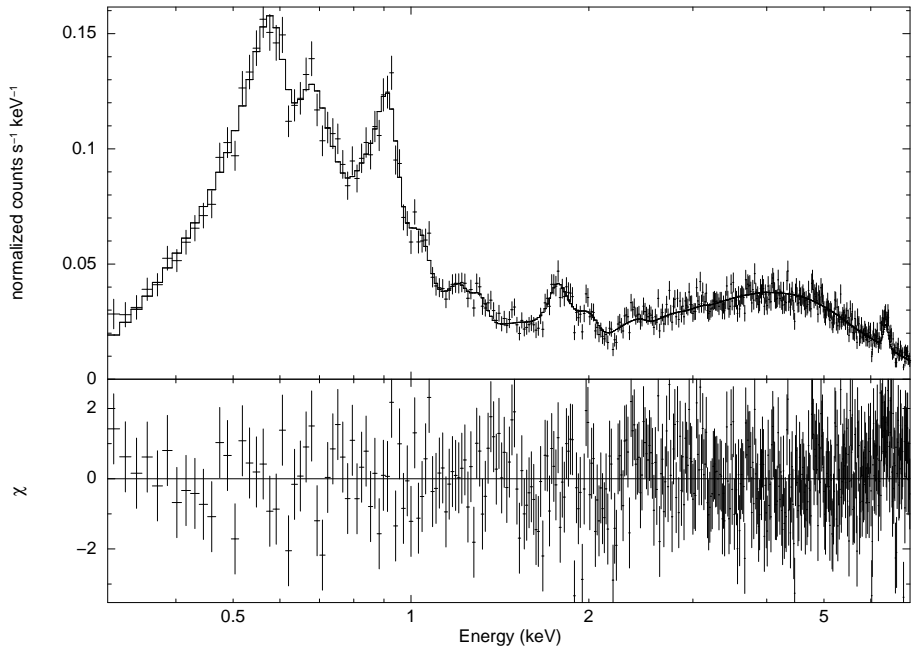}
\caption{Spectrum of the extended X-ray emission with the best-fit
  phenomenological model consisting of a Bremsstrahlung component and
  gaussian lines, with lower panel showing the bin-to-bin contribution to the $\chi^2$ statistics.
\label{4reg2}}
\end{figure}

\begin{figure}
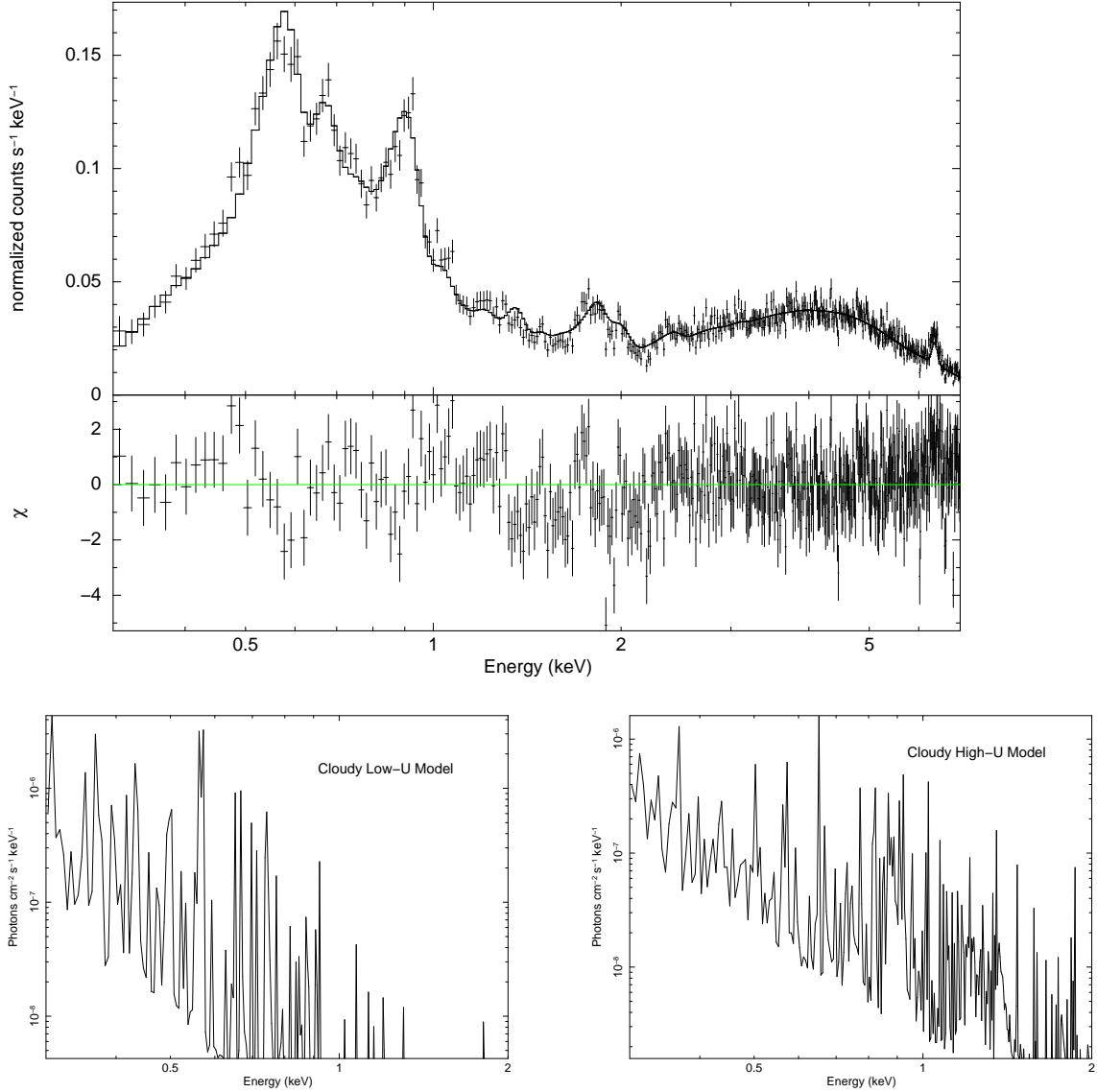

\includegraphics[scale=.55,angle=-90]{f8a.eps}
\includegraphics[scale=.30,angle=-90]{f8b.eps}
\includegraphics[scale=.30,angle=-90]{f8c.eps}
\caption{(a) The same spectrum as shown in Figure~\ref{4reg2}, but
  with the best fit spectral model consisting of two photoionized
  components. (b)-(c) The photoionized models calculated by {\tt
    Cloudy} for the low ionization parameter ($\log U=-0.25$) and the
  high ionization parameter ($\log U=0.8$) component, respectively.
\label{phot3}}
\end{figure}

\clearpage

\begin{figure}
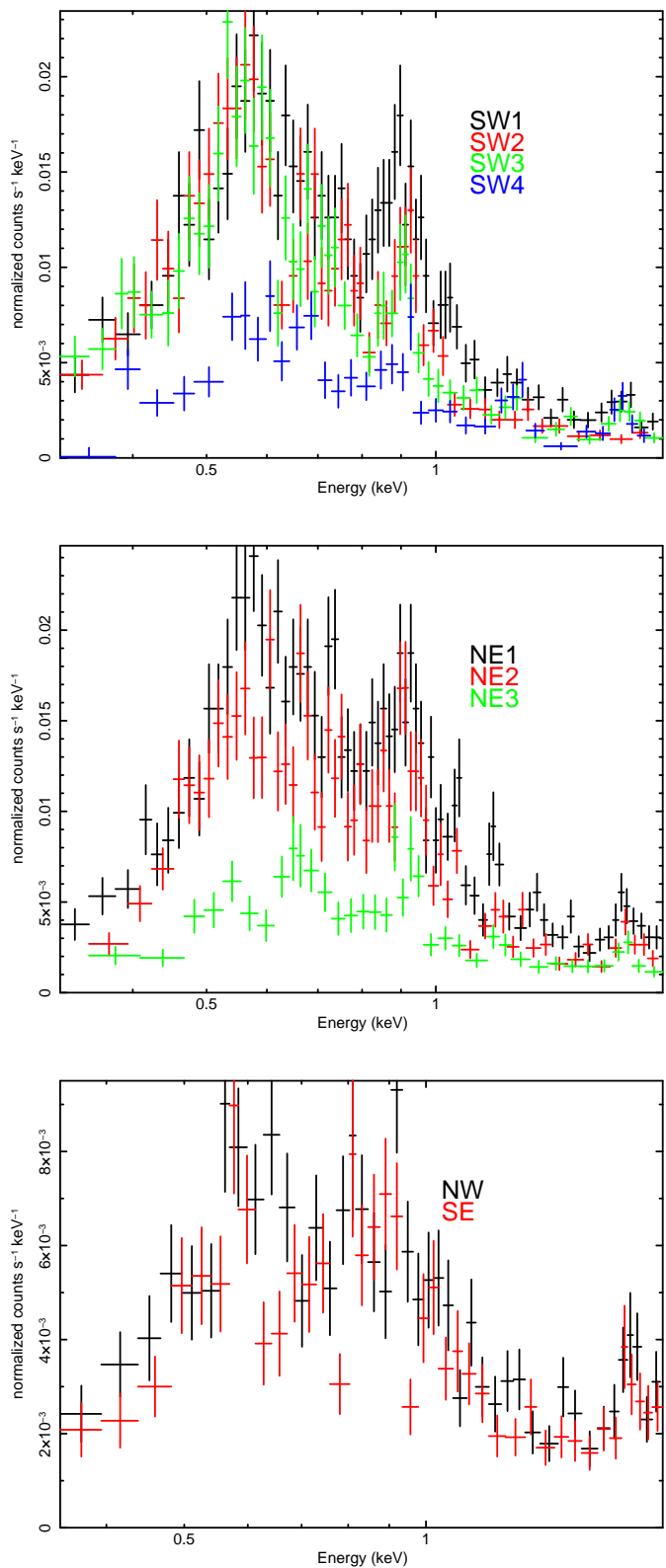

\vspace{-0.3in}
\centering
\includegraphics[scale=.38,angle=-90]{f9a.eps}
\centering
\includegraphics[scale=.38,angle=-90]{f9b.eps}
\centering
\includegraphics[scale=.38,angle=-90]{f9c.eps}
\caption{Comparison of the ACIS spectra across sectors and radially
  binned sub-regions (see Figure~\ref{define}), demonstrating the
  spectral differences. (a) the SW cone; (b) the NE cone; (c) the NW and
  SE sectors that are perpendicular to the bicone.\label{compare4}}
\end{figure}

\clearpage

\begin{figure}
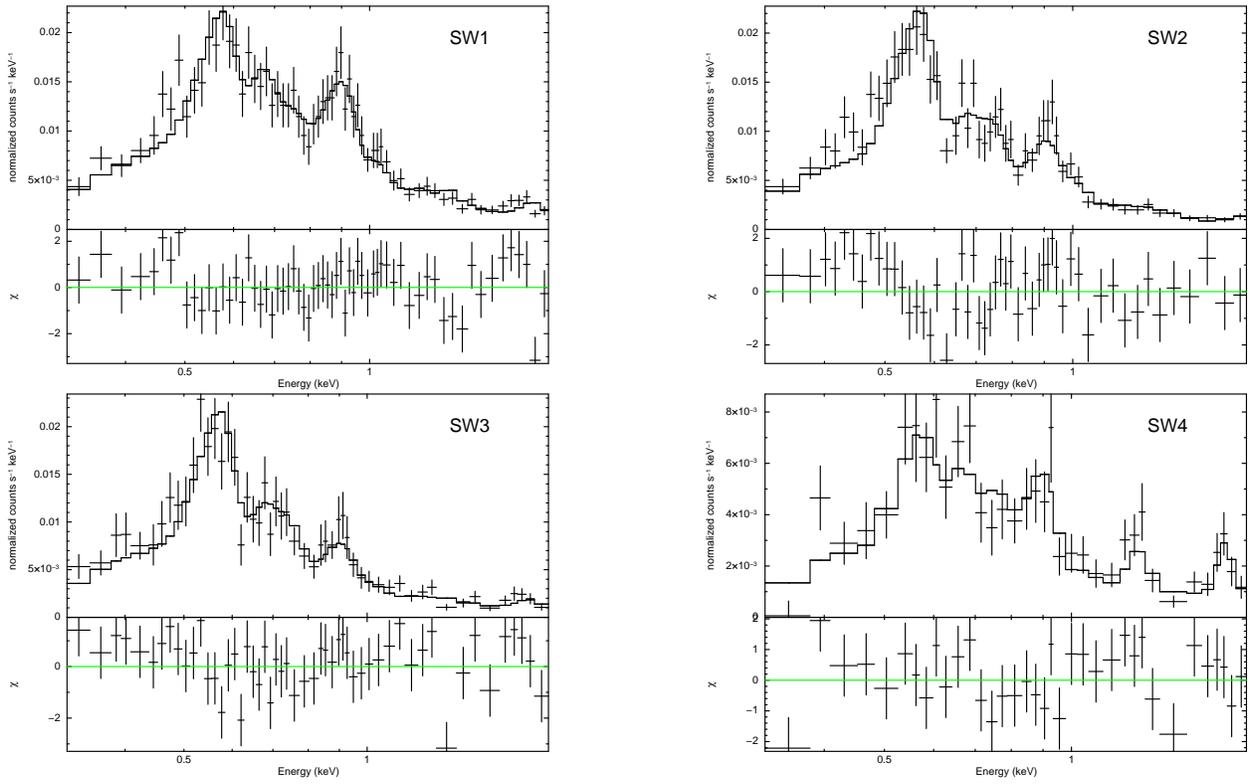

\includegraphics[scale=.3,angle=-90]{f10a.eps}
\includegraphics[scale=.3,angle=-90]{f10b.eps}
\includegraphics[scale=.3,angle=-90]{f10c.eps}
\hspace{0.7in}
\includegraphics[scale=.3,angle=-90]{f10d.eps}
\caption{(a)-(d): Spectra of the X-ray emission in four regions
  radially separated along the SW sector (see Figure~\ref{define}).
\label{sw}}
\end{figure}

\clearpage

\begin{figure}
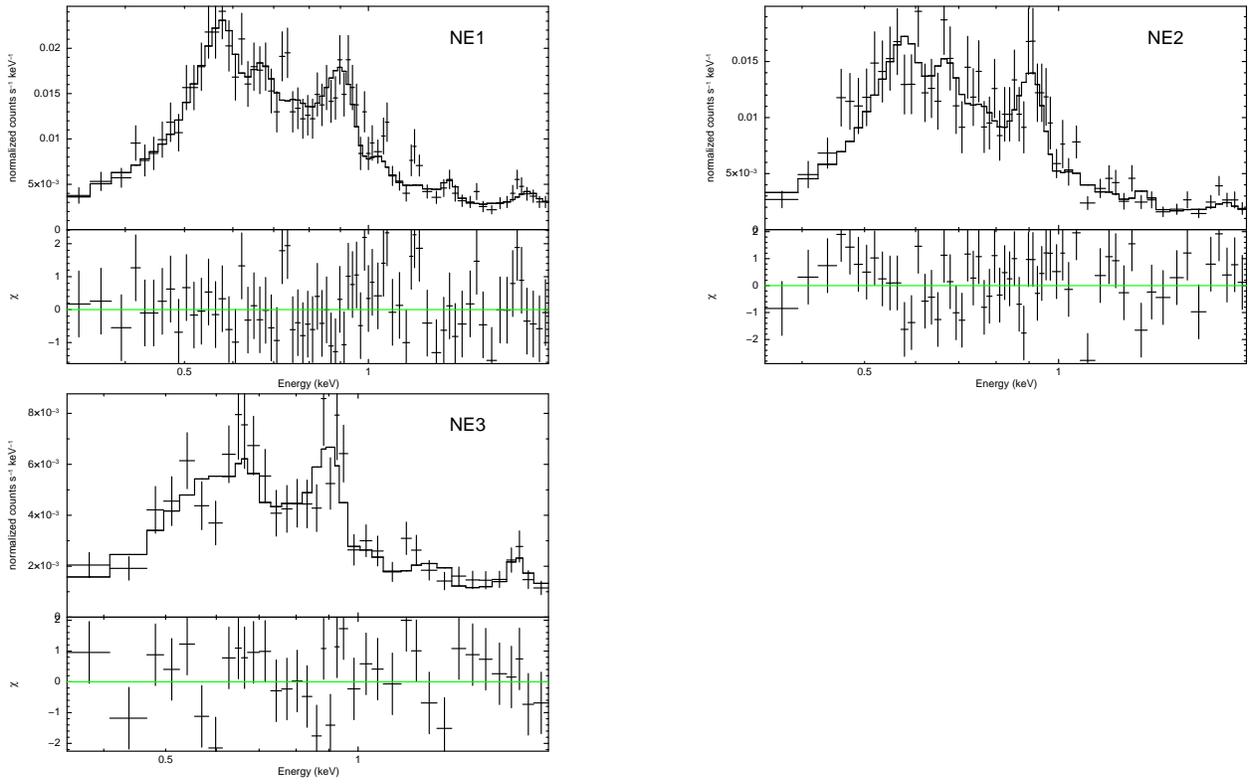

\includegraphics[scale=.30,angle=-90]{f11a.eps}
\includegraphics[scale=.30,angle=-90]{f11b.eps}
\includegraphics[scale=.30,angle=-90]{f11c.eps}
\caption{(a)-(c): Spectra of the X-ray emission in three regions radially along
  the NE sector.
\label{ne}}
\end{figure}

\clearpage

\begin{figure}
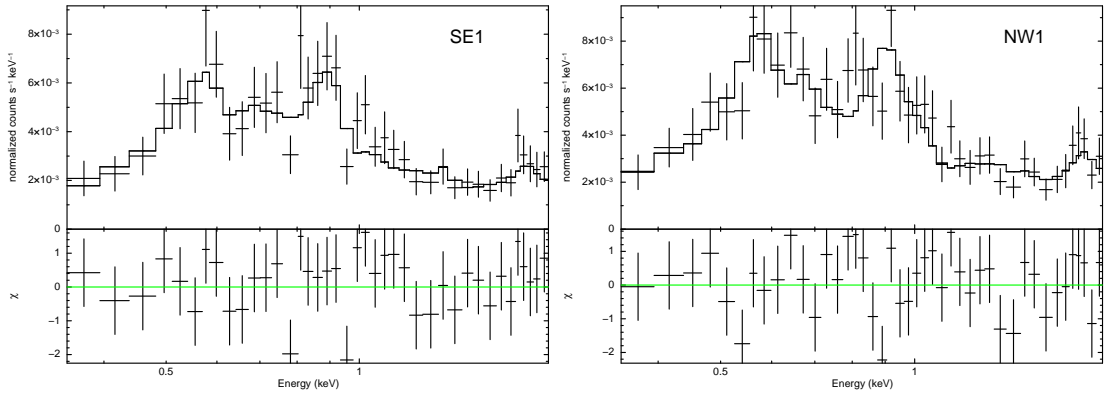

\includegraphics[scale=.30,angle=-90]{f12a.eps}
\includegraphics[scale=.30,angle=-90]{f12b.eps}
\caption{Spectra of the X-ray emission in the SE and NW sectors.
\label{se-nw}}
\end{figure}

\clearpage
\begin{figure}
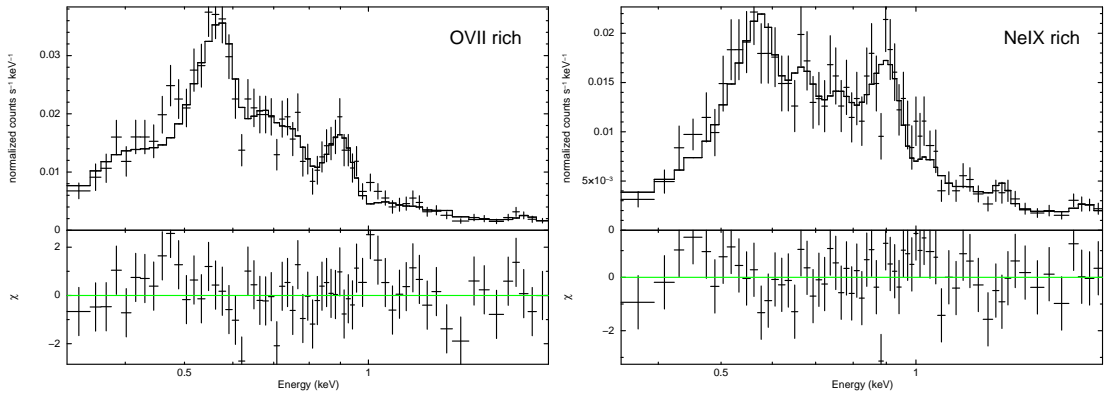

\includegraphics[scale=.30,angle=-90]{f13a.eps}
\includegraphics[scale=.30,angle=-90]{f13b.eps}
\caption{Spectra of the X-ray emission in the bicone regions that are
  prominent in (a) 0.3--0.7 keV emission (red in Figure~\ref{3color})
  and (b) in 0.7--1 keV emission (green in Figure~\ref{3color}).
\label{red-green}}
\end{figure}
\clearpage

\begin{figure}
\epsscale{1.0}
\plotone{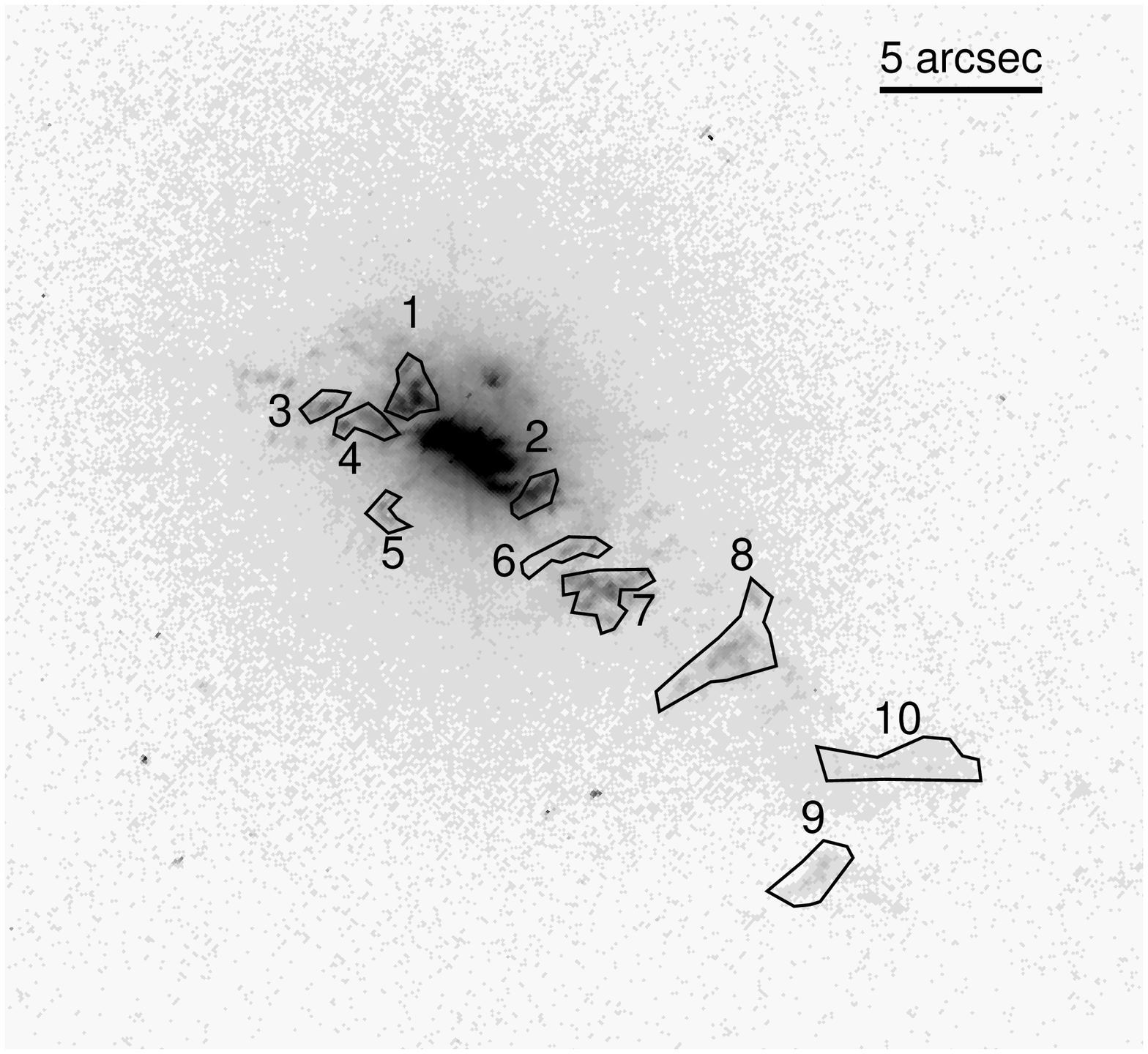}
\caption{HST [OIII] clouds labeled with numbers, for which the
  [OIII]/soft X-ray flux ratio were derived.
\label{oiii_cloud}}
\end{figure}

\clearpage
\begin{figure}
\epsscale{1.0}
\plotone{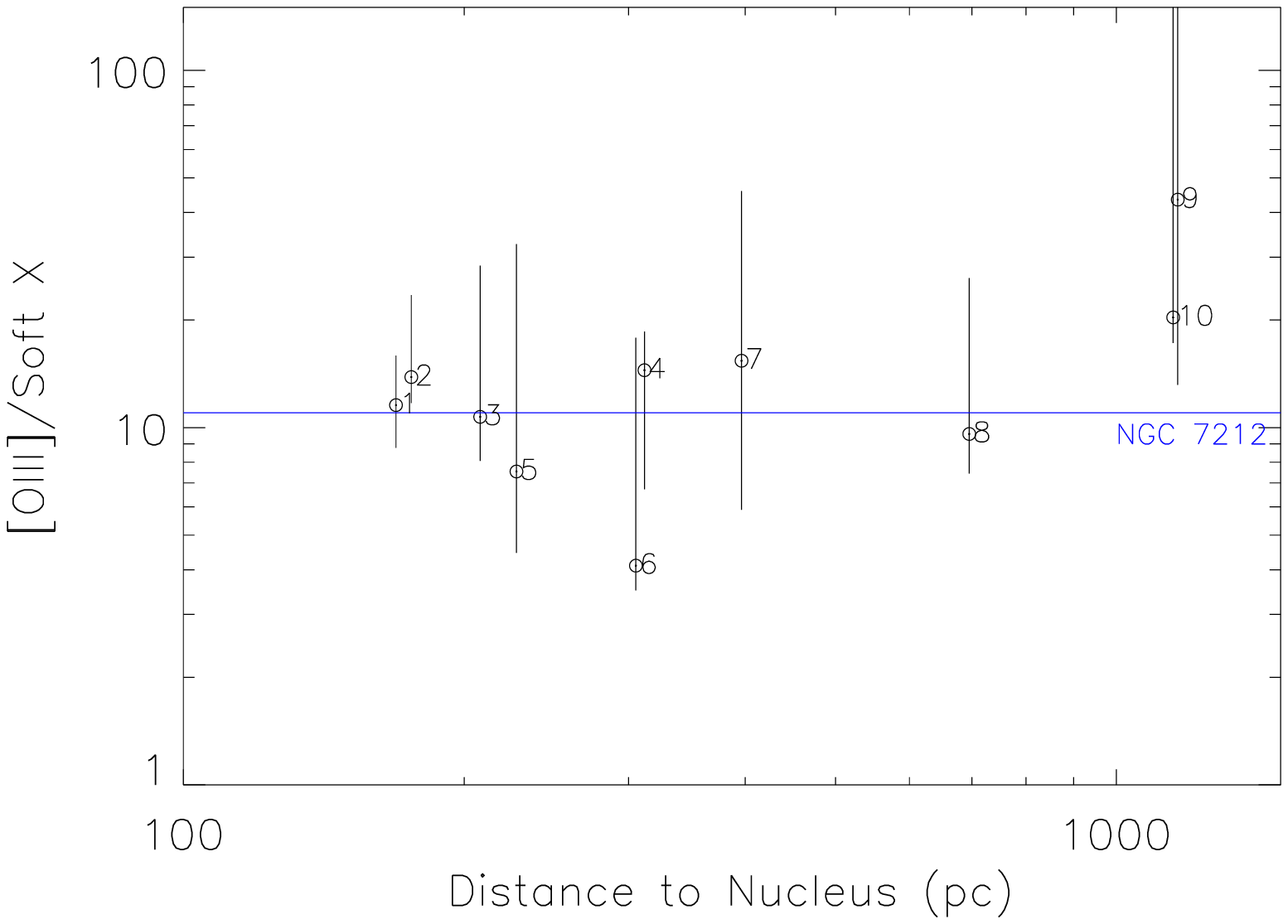}
\caption{Radial dependence of the [OIII]/soft X-ray flux ratio. For comparison, the ratio for the Seyfert 2 galaxy NGC 7212 (Bianchi et al. 2006) is shown.
\label{ratio}}
\end{figure}

\begin{figure}
\epsscale{1.0} 
\plotone{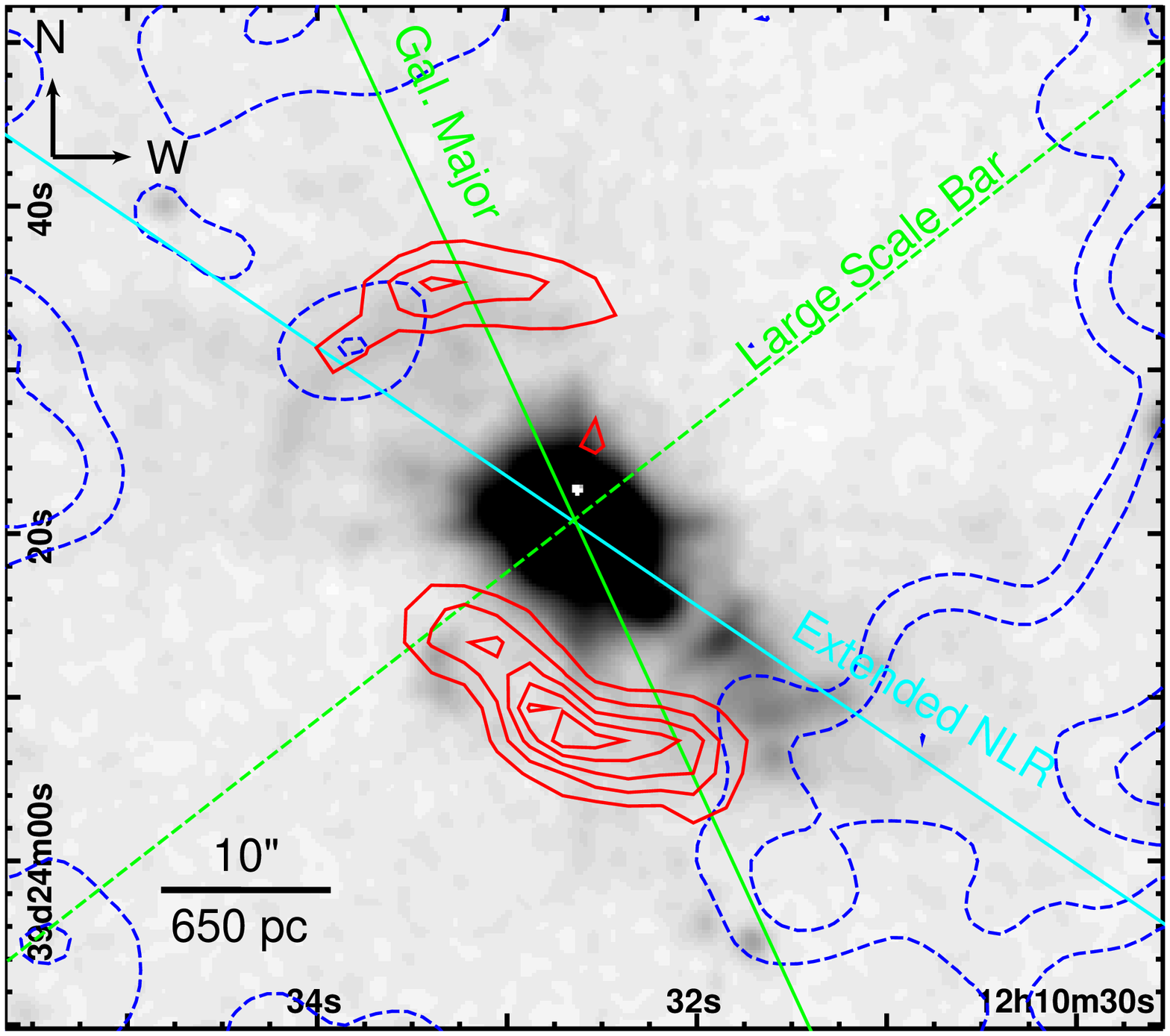} 
\caption{Continuum-subtracted
  H${\alpha}$ image of the central $1\arcmin \times 1\arcmin$ of NGC
  4151 obtained with the Jacobus Kapteyn Telescope (Knapen et
  al. 2004), which illustrates the directions of the kinematic major
  axis of the host galaxy (green solid line; P.A.$\sim$22$^{\circ}$,
  Pedlar et al. 1992; Mundell et al. 1999), the large scale ``weak fat
  bar'' (green dotted line; P.A.$\sim$130$^{\circ}$; Mundell \& Shone
  1999), and the ENLR bicone (cyan line; P.A.$\sim$65$^{\circ}$, Evans
  et al. 1993).  The blue and red contours show the distribution of HI
  (Mundell et al. 1999) and CO (Dumas et al.\ 2010) gas
  respectively.\label{morph}}
\end{figure}

\begin{figure}
\epsscale{0.5}
\plotone{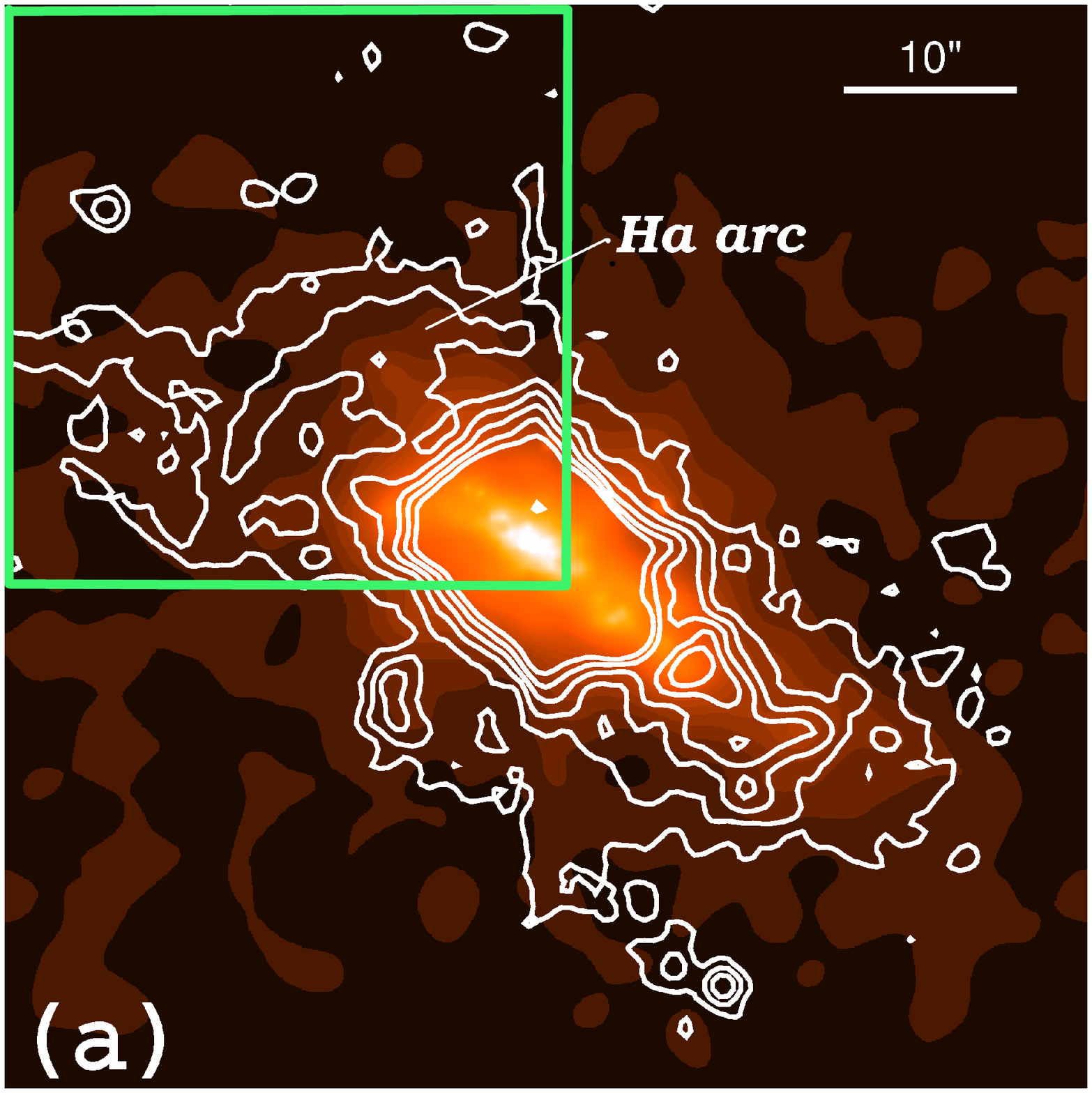}
\epsscale{0.55}
\plotone{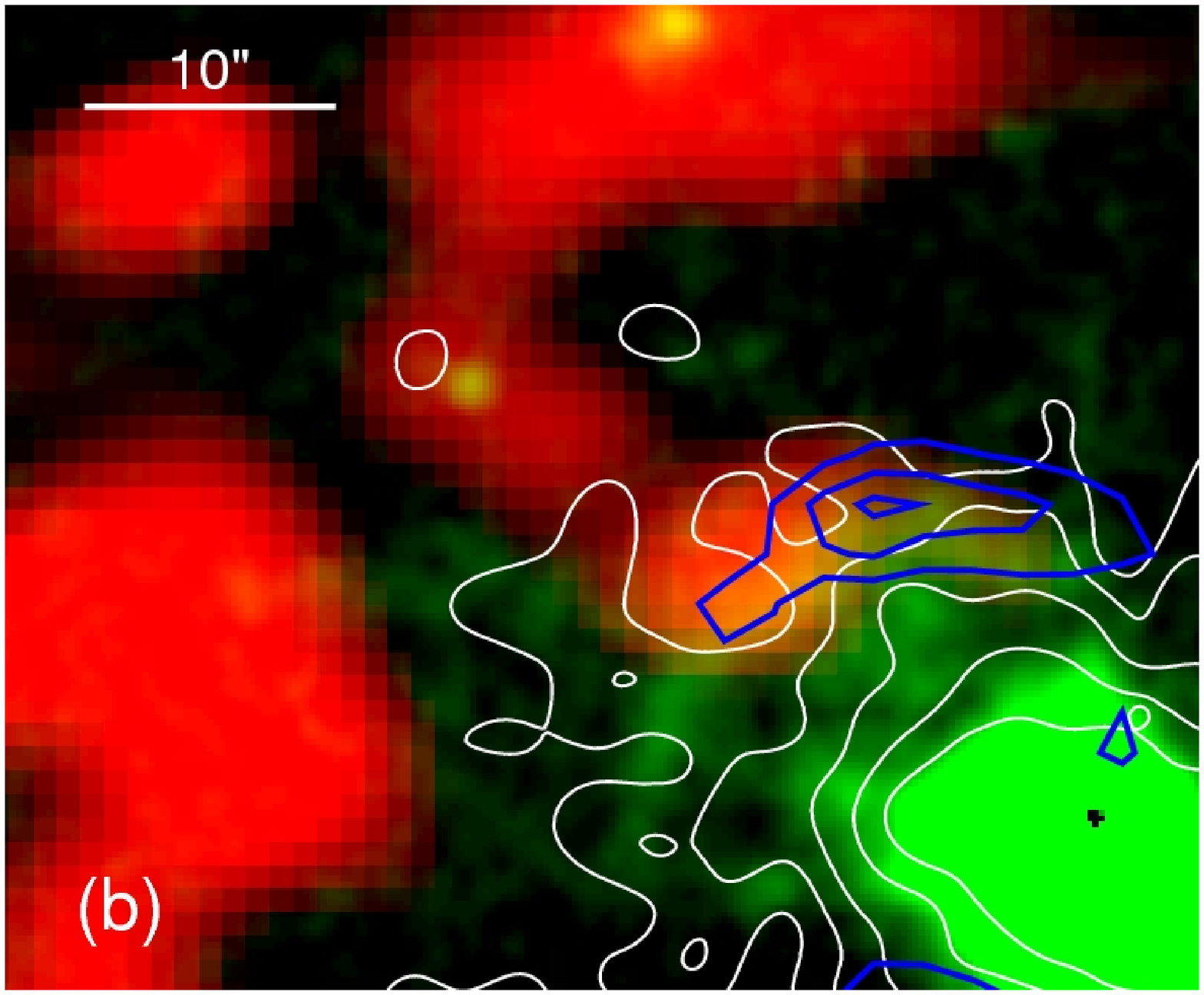}
\epsscale{1.0}
\caption{(a) Superposition of a continuum-subtracted H$\alpha$ image
  (contours; from Knapen et al. 2004) on the central $1\arcmin \times
  1\arcmin$ of the smoothed 0.3--1 keV ACIS image. The box region is
  enlarged in panel $b$. (b) A composite image showing the relative
  distribution of HI emission in red and H$\alpha$ emission in green.
  Overlaid white contours are the soft X-ray emission, and blue
  contours the $^{12}$CO emission.
\label{fig8}}
\end{figure}

\clearpage
\begin{figure}
\epsscale{0.7}
\plotone{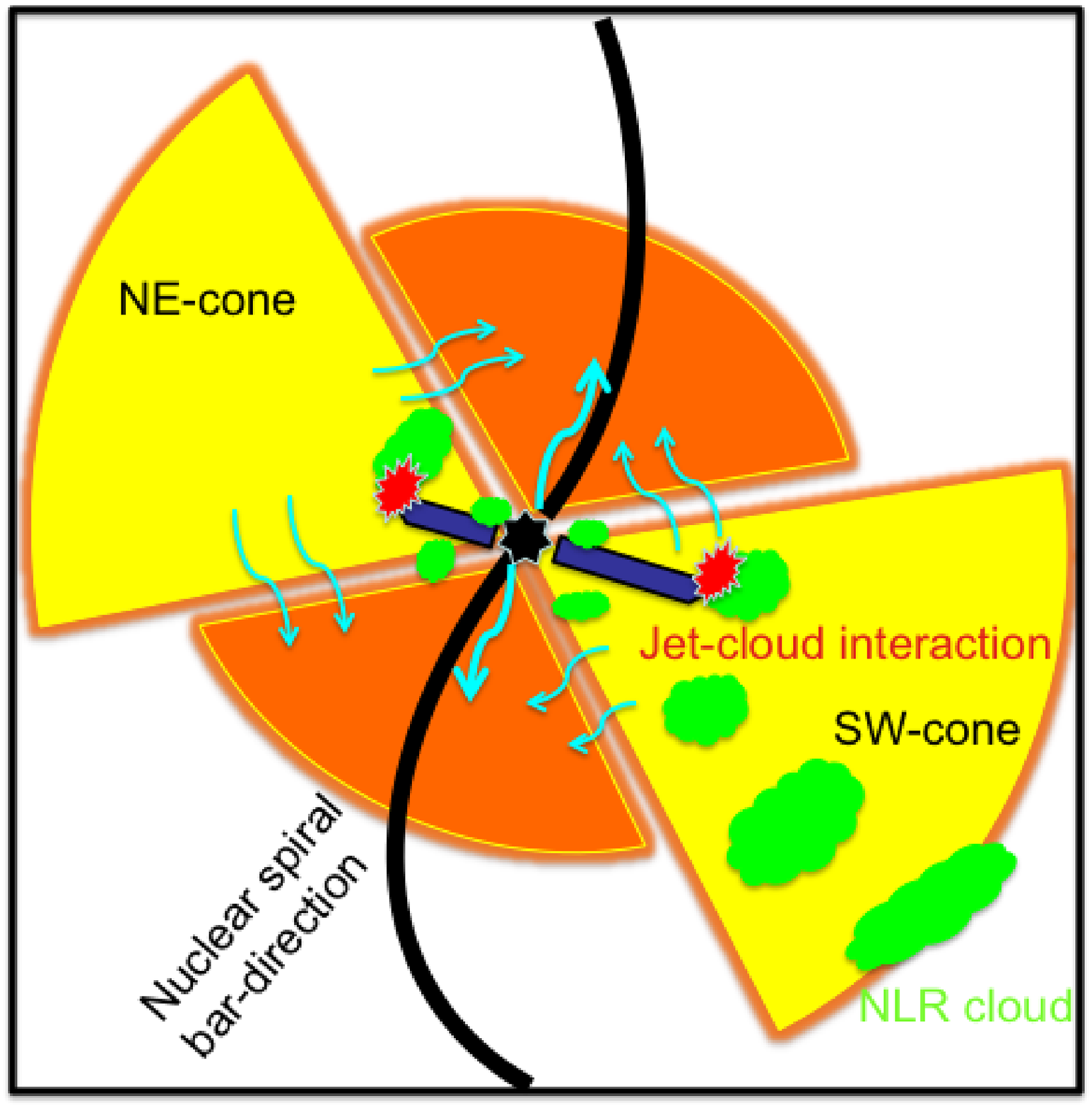}
\plotone{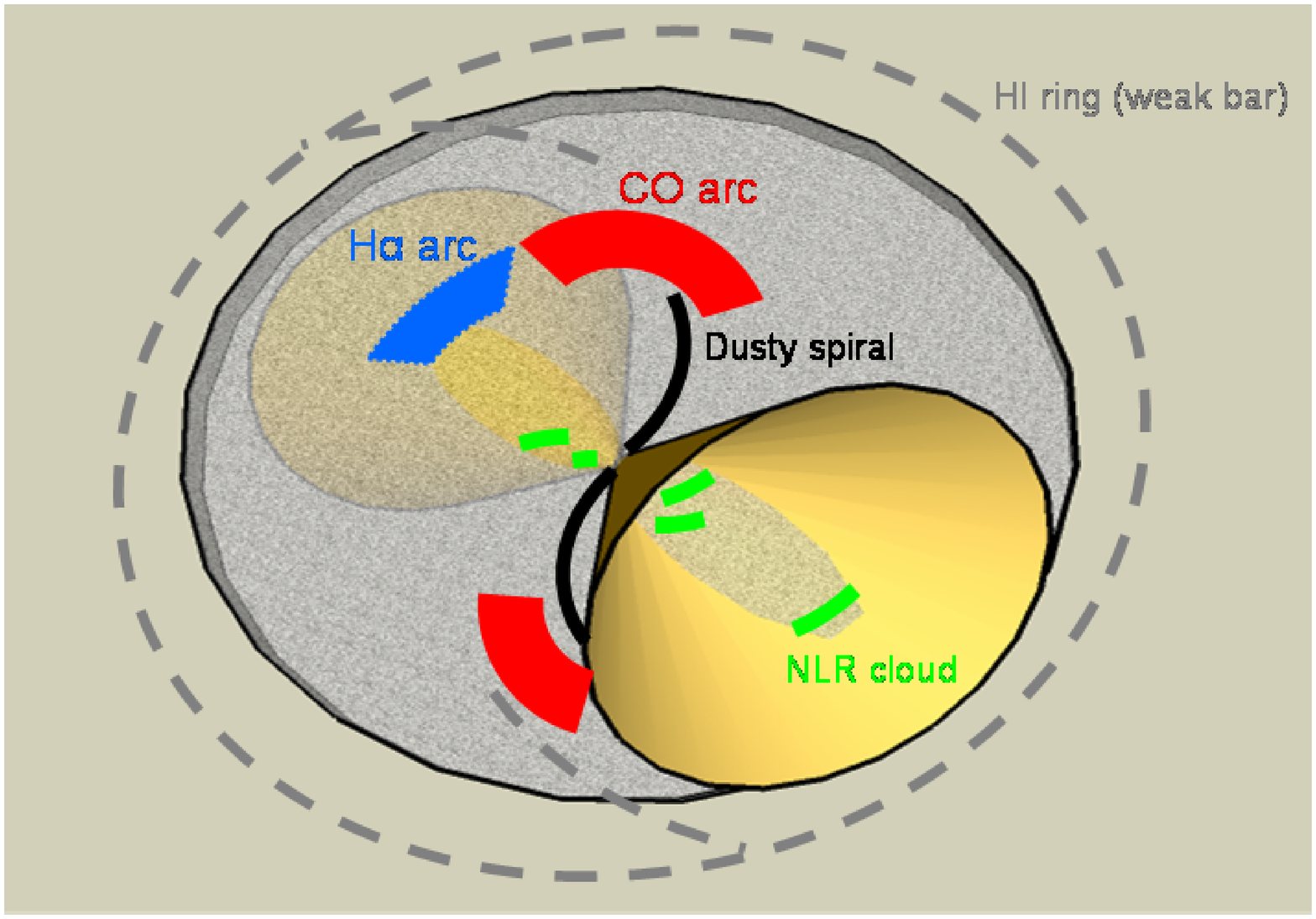}
\caption{A schematic drawing of the complex circum-nuclear environment
  of NGC 4151. (a) The inner few hundred pc radius region; (b) The
  3~kpc-across region.  The features are not drawn to exact spatial
  scale.  The cyan wiggly lines represent possible leakage/scattering
  of ionizing photons, whereas the orange wedges denote the lower
  ionization sectors perpendicular to the bi-cone (yellow wedges).
\label{cartoon}}
\end{figure}

\clearpage
\begin{deluxetable}{cccccc}
\tabletypesize{\scriptsize}
%\tablenum{1}
%\footnotesize       % 10 pt
%\scriptsize           8 pt
\tablewidth{0pt}
%\rotate
\tablecaption{Blended Emission Lines Identified in the Large Scale Extended
  X-ray Emission \label{tab:extended_flux}}
\tablecolumns{6} 
\tablehead{\colhead{ACIS} &\colhead{HETG} &\colhead{Observed Energy\tablenotemark{a}} & \colhead{Restframe Energy} &  \colhead{Line Flux\tablenotemark{b}} & \colhead{ACIS/HETG Flux Ratio\tablenotemark{c}}\\ 
  \colhead{Blend} &  \colhead{Lines} & \colhead{[keV]} & \colhead{[keV]} & \colhead{[photon cm$^{-2}$ s$^{-1}$]}
} 
\startdata
OVII  & & 0.57 & ... & $5.5\times 10^{-5}$ & 13\% \\
...  & OVII f & 0.5617 & 0.5611 & $3.1 \times 10^{-4}$& \\
...  & OVII i & 0.5691 & 0.5687 & $5.1 \times 10^{-5}$& \\
...  & OVII r & 0.5745 & 0.5740 & $6.8 \times 10^{-5}$& \\
\hline \\
OVIII  & & 0.66 & ... & $1.7\times 10^{-5}$ & 13\% \\
...  & OVIII Ly$\alpha$ & 0.6541 & 0.6537 & $1.0\times 10^{-4}$&  \\
...  & NVII RRC & 0.67 & ... & $2.6\times 10^{-5}$& \\
\hline \\
OVII & & 0.73 & ... & $8.5\times 10^{-6}$ & 12\% \\
...  & OVII RRC & 0.74 & ... & $5.2\times 10^{-5}$&  \\
...  & OVIII Ly$\beta$ & 0.7749 & 0.7747 & $1.8\times 10^{-5}$& \\
\hline \\
OVIII & & 0.82 & ... & $4.4\times 10^{-6}$ & 2\% \\
...  & OVIII Ly$\gamma$ & 0.8175 & 0.8171 & $1.1\times 10^{-5}$& \\
...  & OVIII RRC & 0.86 & ... & $1.1\times 10^{-5}$& \\
\hline \\
NeIX & & 0.91 & ... & $1.3\times 10^{-5}$ & 21\% \\
...  & NeIX f & 0.9064 & 0.9052 & $3.3\times 10^{-5}$& \\
... &  NeIX i & 0.9172 & 0.9151 & $9.9\times 10^{-6}$& \\
...  & NeIX r & 0.9229 & 0.9221 & $2.0\times 10^{-5}$& \\
\hline \\
NeX &  &1.03 & ... & $4.4\times 10^{-6}$& 17\% \\
...  & NeX Ly$\alpha$ & 1.0224 & 1.0219 & $2.0\times 10^{-5}$ & \\
...  & NeIX $1s3p-1s^2$ & 1.0749 & 1.0738 & $5.9\times 10^{-6}$& \\
\hline \\
NeX &  &1.20 & ... & $1.9\times 10^{-6}$& 12\% \\
... & NeIX RRC & 1.20 & ... & $9.3\times 10^{-6}$& \\
... & NeX Ly$\beta$ & 1.2118 & 1.2110 & $7.4\times 10^{-6}$& \\
\hline \\
MgXI & & 1.30 & ... & $1.9\times 10^{-6}$ & 9\% \\
... & MgXI f & 1.3542 & 1.3313  & $1.2\times 10^{-5}$& \\
...  & MgXI r & 1.3525 & 1.3524  & $9.3\times 10^{-6}$& \\
\hline \\
SiI &  & 1.75 & ... & $1.6\times 10^{-6}$& 9\% \\
... & SiI K$\alpha$ & 1.7413 & 1.7391  & $1.3\times 10^{-5}$& \\
...  & MgXII Ly$\beta$  & 1.7472 & 1.7450  & $5.3\times 10^{-6}$& \\
\enddata

\tablecomments{The continuum is modeled with a bremsstrahlung (best
  fit $kT=0.31$ keV) absorbed by the Galactic column $N_H=2.1\times
  10^{20}$ cm$^{-2}$ (Murphy et al. 1993).}

\tablenotetext{a}{Emission line energy measured in our ACIS spectrum and in Ogle et al.\ (2000) HETGS spectrum.}
\tablenotetext{b}{Line flux from the present data and in Ogle et al.\ (2000) HETGS spectrum.}
\tablenotetext{c}{Ratio of the ACIS (present data) to HETGS  line fluxes.}
\end{deluxetable}

\clearpage
\pagestyle{empty}
\begin{deluxetable}{cccccccccc}
\tabletypesize{\scriptsize}
%\tablenum{1}
\tablewidth{0pt}
\rotate
\tablecaption{Spectral Fits to the Extended X-ray Emission \label{tab:cloudy}}
\tablecolumns{10} 
\tablehead{\colhead{Region} & \colhead{Source} & \colhead{$\log N_{H,l.o.s}$\tablenotemark{a}} & \colhead{$\log U_1$} & \colhead{$\log N_{H,1}$} & \colhead{$\log U_2$} & \colhead{$\log N_{H,2}$} & \colhead{Nuclear PSF\tablenotemark{b}} & \colhead{$\chi^2$ / d.o.f.} & \colhead{$F_{0.5-2keV}$}\\ 
\colhead{} & \colhead{Counts [0.3--2 keV]} & \colhead{[$\times 10^{22}$ cm$^{-2}$]} & \colhead{}& \colhead{[cm$^{-2}$]} & \colhead{} & \colhead{[cm$^{-2}$]}  & \colhead{Counts [0.3--2 keV]} & \colhead{} & \colhead{erg s$^{-1}$ cm$^{-2}$ arcsec$^{-2}$}}

\startdata
NE  1 & 2432 & 0.04$\pm 0.02$ & 0.86 & 19.7 & -0.22 & 19.7 & $380$  & 63/65  & $6.9\times 10^{-15}$ \\
.... 2 & 1814 & 0.07$\pm 0.03$ & 1.05 & 19.8 & 0.06 & 20.1 & $200$  & 60/53  & $1.3\times 10^{-15}$ \\
.... 3 & 913 & 0.03$\pm 0.01$ & 0.49 & 19.9 & -1.87 & 23.5 & $140$  & 26/29  & $1.7\times 10^{-16}$\\
SW  1 & 2096 & 0.02  & 0.83 & 20.0 & -0.38 & 19.4 & $270$  & 60/57  & $1.0\times 10^{-14}$ \\
....2 & 1667 & 0.02 & 1.06 & 20.0 & -0.42 & 19.8 & $130$  & 58/45 & $2.9\times 10^{-15}$ \\
....3 & 1560 & 0.02 & 1.00 & 19.6 & -0.50 & 19.6 & $100$   & 54/46  & $7.2\times 10^{-16}$ \\
....4 & 966 & 0.02 & 0.34 & 19.0 & -1.28 & 23.5 & $80$   &  34/30 & $4.5\times 10^{-17}$\\
NW  & 1050 & 0.22$\pm 0.02$ & 1.03 & 19.2 & -0.98 & 19.5 & $400$ &  35/35 &  $3.1\times 10^{-16}$ \\
SE  & 1081 & 0.16$\pm 0.04$ & 1.26 & 20.3 & -1.16 & 19.9 & $530$ & 28/33  & $2.2\times 10^{-16}$ \\
\hline
OVII-rich & 2626 & 0.02$\pm 0.02$ & 0.14 & 20.3  &  -1.22  & 20.3 & $200$ & 60/56  & $4.3\times 10^{-15}$ \\
NeIX-rich  &  2141 & 0.07$\pm 0.02$ & 0.6 & 19.6 & -0.21  & 20.4  & $210$ & 50/56  & $4.0\times 10^{-15}$ \\
\enddata

\tablenotetext{a}{The line-of-sight absorbing column density towards
  the photoionized emission. For the SW regions, the fitted
  $N_{H,l.o.s}$ prefers values lower than the Galactic column towards
  NGC 4151, and was fixed at $N_H=2\times 10^{20}$ \citep{Murphy96}.}

\tablenotetext{b}{Expected counts from simulation of PSF scattered
  nuclear emission.}

\end{deluxetable}

\clearpage

\begin{deluxetable}{cccccc}
%\rotate
\tabletypesize{\scriptsize}
%\tabletypesize{\small}
\tablecaption{Measured X-ray and [OIII] Fluxes for the Clouds \label{fluxratio}}
\tablewidth{0pt}
\tablehead{
\colhead{\begin{tabular}{c}
Cloud\\
Number\\
\end{tabular}} &
\colhead{\begin{tabular}{c}
Distance\\
to Nuc. ($\arcsec$)\\
\end{tabular}} &
\colhead{\begin{tabular}{c}
Distance\\
to Nuc. (pc)\\
\end{tabular}} &
\colhead{\begin{tabular}{c}
[OIII] flux\\
($10^{-14}$ergs s$^{-1}$ cm$^{-2}$)\\
\end{tabular}} &
\colhead{\begin{tabular}{c}
0.5-2 keV flux\\
($10^{-15}$ergs s$^{-1}$ cm$^{-2}$)\\
\end{tabular}} &
\colhead{\begin{tabular}{c}
$F_{[OIII]}/F_{[0.5-2 keV]}$\\
\end{tabular}}
}
\startdata
1 & 2.6 & 169 & 29.8$\pm 1.9$ & 25.7$\pm 5.2$ &  11.6\\
2 & 2.7 & 175 & 21.1$\pm 2.0$ & 15.2$^{+1.0}_{-5.4}$ & 13.8\\
3 & 3.2 & 208 & 2.6$\pm 0.4$ & 2.4$^{+0.3}_{-1.3}$  & 10.7\\
4 & 4.8 & 312 & 7.9$\pm 0.9$ & 5.4$^{+5.0}_{-0.7}$  & 14.5 \\
5 & 3.5 & 228 & 1.2$\pm 0.4$ & 1.6$^{+0.2}_{-1.0}$  & 7.5 \\
6 & 4.7 & 306 & 1.9$\pm 0.2$ & 4.7$^{+0.2}_{-3.5}$  & 4.1 \\
7 & 6.1 & 397 & 10.2$\pm 6.0$ & 6.6$^{+0.5}_{-3.1}$  & 15.2\\
8 & 10.7 & 696 & 3.9$\pm 0.7$ & 4.2$^{+0.2}_{-2.5}$  & 9.6\\
9 & 17.9 & 1164 & 0.9$\pm 0.1$ & 0.2$^{+0.4}_{-0.2}$  & 43.4 \\
10 & 17.7 & 1151 & 1.8$\pm 0.1$ & 0.9$^{+0.1}_{-0.8}$  & 20.3 \\
\enddata
\tablecomments{The [OIII] flux is measured from F502N image.}
\end{deluxetable}


\begin{thebibliography}{}



\bibitem[Armentrout et al.(2007)]{2007ApJ...665..237A} Armentrout, B.~K., Kraemer, S.~B., \& Turner, T.~J.\ 2007, \apj, 665, 237 

\bibitem[Arnaud(1996)]{1996ASPC..101...17A} Arnaud, K.~A.\ 1996, Astronomical Data Analysis Software and Systems V, 101, 17 

\bibitem[Asif et al.(1998)]{Asif98} Asif, M.~W., Mundell, C.~G., Pedlar, A., Unger, S.~W., Robinson, A., Vila-Vilaro, B., \& Lewis, J.~R.\ 1998, \aap, 333, 466

\bibitem[Asif et al.(2005)]{2005MNRAS.359..408A} Asif, M.~W., Mundell, C.~G., \& Pedlar, A.\ 2005, \mnras, 359, 408 

\bibitem[Barbosa et al.(2009)]{2009MNRAS.396....2B} Barbosa, F.~K.~B., Storchi-Bergmann, T., Cid Fernandes, R., Winge, C., \& Schmitt, H.\ 2009, \mnras, 396, 2 

\bibitem[Bentz et al.(2006)]{2006ApJ...651..775B} Bentz, M.~C., et al.\ 2006, \apj, 651, 775 

\bibitem[Bianchi et al.(2006)]{2006A&A...448..499B} Bianchi, S., Guainazzi, M., \& Chiaberge, M.\ 2006, \aap, 448, 499 

\bibitem[Bianchi et al.(2010)]{2010MNRAS.405..553B} Bianchi, S., Chiaberge, M., Evans, D.~A., Guainazzi, M., Baldi, R.~D., Matt, G., \& Piconcelli, E.\ 2010, \mnras, 405, 553 

\bibitem[Crenshaw et al.(2000)]{2000AJ....120.1731C} Crenshaw, D.~M., et al.\ 2000, \aj, 120, 1731


\bibitem[Crenshaw \& Kraemer(2007)]{2007ApJ...659..250C} Crenshaw, D.~M., \& Kraemer, S.~B.\ 2007, \apj, 659, 250 

\bibitem[Crenshaw et al.(2010)]{2010AJ....139..871C} Crenshaw, D.~M., Kraemer, S.~B., Schmitt, H.~R., Jaff{\'e}, Y.~L., Deo, R.~P., Collins, N.~R., \& Fischer, T.~C.\ 2010, \aj, 139, 871 

\bibitem[Das et al.(2005)]{2005AJ....130..945D} Das, V., et al.\ 2005, \aj, 130, 945

\bibitem[de Rosa et al.(2007)]{2007A&A...463..903D} de Rosa, A., Piro, L., Perola, G.~C., Capalbi, M., Cappi, M., Grandi, P., Maraschi, L., \& Petrucci, P.~O.\ 2007, \aap, 463, 903 

\bibitem[Dumas et al.(2010)]{Dumas10} Dumas, G., Schinnerer, E., \& Mundell, C.~G.\ 2010, \apj, in press, astro-ph/1008.0856

\bibitem[Elitzur \& Shlosman(2006)]{Elitzur06} Elitzur, M., \& Shlosman, I.\ 2006, \apjl, 648, L101 

\bibitem[Elvis et al.(1983)]{1983ApJ...268..105E} Elvis, M., Briel, U.~G., \& Henry, J.~P.\ 1983, \apj, 268, 105 

\bibitem[Elvis et al.(1990)]{1990ApJ...361..459E} Elvis, M., Fassnacht, C., Wilson, A.~S., \& Briel, U.\ 1990, \apj, 361, 459 

\bibitem[Evans et al.(1993)]{1993ApJ...417...82E} Evans, I.~N., Tsvetanov, Z., Kriss, G.~A., Ford, H.~C., Caganoff, S., \& Koratkar, A.~P.\ 1993, \apj, 417, 82 

\bibitem[Ferland et al.(1998)]{1998PASP..110..761F} Ferland, G.~J., Korista, K.~T., Verner, D.~A., Ferguson, J.~W., Kingdon, J.~B., \& Verner, E.~M.\ 1998, \pasp, 110, 761

\bibitem[Fischer et al.(2011)]{2011ApJ...727...71F} Fischer, T.~C., Crenshaw, D.~M., Kraemer, S.~B., Schmitt, H.~R., Mushotsky, R.~F., \& Dunn, J.~P.\ 2011, \apj, 727, 71 

\bibitem[Garmire et al.(2003)]{2003SPIE.4851...28G} Garmire, G.~P., Bautz, M.~W., Ford, P.~G., Nousek, J.~A., \& Ricker, G.~R., Jr.\ 2003, \procspie, 4851, 28

\bibitem[Gonz{\'a}lez-Mart{\'{\i}}n (2008)]{Omai08} Gonz{\'a}lez-Mart{\'{\i}}n, O.,\ 2008, PhD Thesis, Instituto de  Astrof{\'{\i}}sica de Andaluc{\'{\i}}a, Granada, Spain

\bibitem[Gonzalez-Martin et al.(2010)]{2010ApJ...723.1748G} Gonzalez-Martin, O., Acosta-Pulido, J.~A., Perez Garcia, A.~M., \& Ramos Almeida, C.\ 2010, \apj, 723, 1748 

\bibitem[Guainazzi et al.(2009)]{2009A&A...505..589G} Guainazzi, M., Risaliti, G., Nucita, A., Wang, J., Bianchi, S., Soria, R., \& Zezas, A.\ 2009, \aap, 505, 589 

\bibitem[Heckman \& Balick(1983)]{1983ApJ...268..102H} Heckman, T.~M., \& Balick, B.\ 1983, \apj, 268, 102 

\bibitem[Holt et al.(2006)]{2006MNRAS.370.1633H} Holt, J., Tadhunter, C., Morganti, R., Bellamy, M., Gonz{\'a}lez Delgado, R.~M., Tzioumis, A., \& Inskip, K.~J.\ 2006, \mnras, 370, 1633 

\bibitem[Holt et al.(2011)]{2011MNRAS.410.1527H} Holt, J., Tadhunter, C.~N., Morganti, R., \& Emonts, B.~H.~C.\ 2011, \mnras, 410, 1527 

\bibitem[Hopkins \& Elvis(2010)]{Hopkins10} Hopkins, P.~F., \& Elvis, M.\ 2010, \mnras, 401, 7 

\bibitem[Kaiser et al.(2000)]{Kaiser00} Kaiser, M.~E., et al.\ 2000, \apj, 528, 260

\bibitem[Kaspi et al.(2005)]{2005ApJ...629...61K} Kaspi, S., Maoz, D., Netzer, H., Peterson, B.~M., Vestergaard, M., \& Jannuzi, B.~T.\ 2005, \apj, 629, 61 

\bibitem[Kinkhabwala et al.(2002)]{2002ApJ...575..732K} Kinkhabwala, A., et 
al.\ 2002, \apj, 575, 732 

\bibitem[Kinkhabwala et al.(2003)]{2003astro.ph..4332K} Kinkhabwala, A., 
Behar, E., Sako, M., Gu, M.~F., Kahn, S.~M., 
\& Paerels, F.~B.~S.\ 2003, arXiv:astro-ph/0304332 

\bibitem[Knapen et al.(2004)]{2004A&A...426.1135K} Knapen, J.~H., Stedman, S., Bramich, D.~M., Folkes, S.~L., \& Bradley, T.~R.\ 2004, \aap, 426, 1135 

\bibitem[Komossa(2001)]{2001A&A...371..507K} Komossa, S.\ 2001, \aap, 371, 507 

\bibitem[Kraemer et al.(2005)]{Kraemer05} Kraemer, S.~B., et al.\  2005, \apj, 633, 693 

\bibitem[Kraemer et al.(2008)]{Kraemer08} Kraemer, S.~B., Schmitt, H.~R., \& Crenshaw, D.~M.\ 2008, \apj, 679, 1128

\bibitem[Krongold et al.(2007)]{Krongold07} Krongold, Y., Nicastro, F., Elvis, M., Brickhouse, N., Binette, L., Mathur, S., \& Jim{\'e}nez-Bail{\'o}n, E.\ 2007, \apj, 659, 1022 



\bibitem[Lehnert et al.(1999)]{1999ApJ...523..575L} Lehnert, M.~D., Heckman, T.~M., \& Weaver, K.~A.\ 1999, \apj, 523, 575 

\bibitem[Liedahl(1999)]{1999LNP...520..189L} Liedahl, D.~A.\ 1999, X-Ray Spectroscopy in Astrophysics, 520, 189 

\bibitem[Lubinski et al.(2010)]{2010arXiv1005.0842L} Lubinski, P., Zdziarski, A.~A., Walter, R., Paltani, S., Beckmann, V., Soldi, S., Ferrigno, C., \& Courvoisier, T.~J.~-.\ 2010, arXiv:1005.0842 

\bibitem[Mathur et al.(2009)]{2009AIPC.1201...33M} Mathur, S., Stoll, R., Krongold, Y., Nicastro, F., Brickhouse, N.,  \& Elvis, M.\ 2009, American Institute of Physics Conference Series, 1201, 33 

\bibitem[Morse et al.(1995)]{1995ApJ...439..121M} Morse, J.~A., Wilson, A.~S., Elvis, M., \& Weaver, K.~A.\ 1995, \apj, 439, 121 

\bibitem[Murphy et al.(1996)]{Murphy96} Murphy, E.~M., Lockman, F.~J., Laor, A., \& Elvis, M.\ 1996, \apjs, 105, 369 

\bibitem[Mundell et al.(2003)]{Mundell03} Mundell, C.~G., Wrobel, J.~M., Pedlar, A., \& Gallimore, J.~F.\ 2003, \apj, 583, 192

\bibitem[Mundell \& Shone(1999)]{MS99} Mundell, C.~G., \& Shone, D.~L.\ 1999, \mnras, 304, 475

\bibitem[Mundell et al.(1999)]{Mundell99} Mundell, C.~G., Pedlar, A., Shone, D.~L., \& Robinson, A.\ 1999, \mnras, 304, 481

\bibitem[Ogle et al.(2000)]{2000ApJ...545L..81O} Ogle, P.~M., Marshall, H.~L., Lee, J.~C., \& Canizares, C.~R.\ 2000, \apjl, 545, L81 

\bibitem[Ogle et al.(2003)]{2003A&A...402..849O} Ogle, P.~M.,
Brookings, T., Canizares, C.~R., Lee, J.~C., \& Marshall, H.~L.\ 2003,
\aap, 402, 849

\bibitem[Osterbrock(1989)]{Oster89} Osterbrock, D.~E.\ 1989,  Astrophysics of gaseous nebulae and active galactic nuclei, Mill Valley, University Science Books



\bibitem[Osterbrock \& Ferland(2006)]{2006agna.book.....O} Osterbrock, D.~E., \& Ferland, G.~J.\ 2006, Astrophysics of gaseous nebulae and active galactic nuclei, 2nd.~ed.~by D.E.~Osterbrock and G.J.~Ferland.~Sausalito, CA: University Science Books, 2006,  



\bibitem[Pedlar et al.(1992)]{1992MNRAS.259..369P} Pedlar, A., Howley, P., Axon, D.~J., \& Unger, S.~W.\ 1992, \mnras, 259, 369 

\bibitem[Penston et al.(1990)]{Penston90} Penston, M.~V., et al.\ 1990, \aap, 236, 53 

\bibitem[Perez et al.(1989)]{1989MNRAS.241P..31P} Perez, E.,  Gonzalez-Delgado, R., Tadhunter, C., \& Tsvetanov, Z.\ 1989, \mnras, 241, 31P 

\bibitem[Porter et al.(2006)]{2006PASP..118..920P} Porter, R.~L., Ferland, G.~J., Kraemer, S.~B., Armentrout, B.~K., Arnaud, K.~A., \& Turner, T.~J.\ 2006, \pasp, 118, 920 

\bibitem[Riffel et al.(2010)]{2010MNRAS.404..166R} Riffel, R.~A., Storchi-Bergmann, T., \& Nagar, N.~M.\ 2010, \mnras, 404, 166

\bibitem[Robinson et al.(1994)]{Robinson94} Robinson, A., et al.\ 1994, \aap, 291, 351

\bibitem[Schnorr M{\"u}ller et al.(2011)]{2011MNRAS.tmp...62S} Schnorr M{\"u}ller, A., Storchi-Bergmann, T., Riffel, R.~A., Ferrari, F., Steiner, J.~E., Axon, D.~J., \& Robinson, A.\ 2011, \mnras, 62 

\bibitem[Schulz \& Komossa(1993)]{1993A&A...278...29S} Schulz, H., \& Komossa, S.\ 1993, \aap, 278, 29 

\bibitem[Schurch et al.(2004)]{2004MNRAS.350....1S} Schurch, N.~J., Warwick, R.~S., Griffiths, R.~E., \& Kahn, S.~M.\ 2004, \mnras, 350, 1 

\bibitem[Shakura \& Sunyaev(1973)]{1973A&A....24..337S} Shakura, N.~I., \& Sunyaev, R.~A.\ 1973, \aap, 24, 337 


\bibitem[Smith et al.(2001)]{2001ApJ...556L..91S} Smith, R.~K., Brickhouse, 
N.~S., Liedahl, D.~A., \& Raymond, J.~C.\ 2001, \apjl, 556, L91 


\bibitem[Strickland et al.(2002)]{2002ApJ...568..689S} Strickland, D.~K., Heckman, T.~M., Weaver, K.~A., Hoopes, C.~G., \& Dahlem, M.\ 2002, \apj, 568, 689 



\bibitem[Storchi-Bergmann et al.(2009)]{SB09} Storchi-Bergmann, T., McGregor, P.~J., Riffel, R.~A., Sim{\~o}es Lopes, R., Beck, T., \& Dopita, M.\ 2009, \mnras, 394, 1148



\bibitem[Storchi-Bergmann et al.(2010)]{SB10}  Storchi-Bergmann, T., Lopes, R.~D.~S., McGregor, P.~J., Riffel, R.~A., Beck, T., \& Martini, P.\ 2010, \mnras, 402, 819 

\bibitem[Unger et al.(1987)]{1987MNRAS.228..671U} Unger, S.~W., Pedlar, A., Axon, D.~J., Whittle, M., Meurs, E.~J.~A., \& Ward, M.~J.\ 1987, \mnras, 228, 671 

\bibitem[Veilleux et al.(2005)]{2005ARA&A..43..769V} Veilleux, S., Cecil, G., \& Bland-Hawthorn, J.\ 2005, \araa, 43, 769 



\bibitem[Wang et al.(2009)]{Wang09b} Wang, J., Fabbiano, G., Karovska, M., Elvis, M., Risaliti, G., Zezas, A., \& Mundell, C.~G.\ 2009, \apj, 704, 1195



\bibitem[Wang et al.(2010a)]{2010ApJ...719L.208W} Wang, J., Fabbiano, G., Risaliti, G., Elvis, M., Mundell, C.~G., Dumas, G., Schinnerer, E., \& Zezas, A.\ 2010a, \apjl, 719, L208 


\bibitem[Wang et al.(2010b)]{Wang10_NUC} Wang, J., Risaliti, G., Fabbiano, G., Elvis, M., Zezas, A., \& Karovska, M.\ 2010b, \apj, 714, 1497

\bibitem[Wang et al.(2011a)]{Wang11a} Wang, J., et al.\ 2011a, \apj, 729, 75 

\bibitem[Wang et al.(2011b)]{Wang2011b} Wang, J., et al.\ 2011b, Submitted to \apj, astro-ph/1103.1912

\bibitem[Weaver et al.(1994)]{1994ApJ...423..621W} Weaver, K.~A., et al.\ 1994, \apj, 423, 621 

\bibitem[Yang et al.(2001)]{Yang01} Yang, Y., Wilson, A.~S., \& Ferruit, P.\ 2001, \apj, 563, 124

\bibitem[Young et al.(2001)]{2001ApJ...556....6Y} Young, A.~J., Wilson, A.~S., \& Shopbell, P.~L.\ 2001, \apj, 556, 6


\end{thebibliography}
\end{document}